\documentclass[sigconf]{acmart}

\usepackage{microtype}\usepackage{enumerate}

\usepackage[normalem]{ulem}
\usepackage{xspace,amstext,url}
\usepackage{colortbl}

\usepackage{mathtools}

\usepackage{algorithm} 
\usepackage[noend]{algpseudocode}

\usepackage[explicit]{titlesec}

\titleformat{\paragraph}[runin]
    {\normalfont\bfseries}
    {}
    {0pt}
    {#1}

\titlespacing{\paragraph}
    {0pt} {1ex plus .1ex minus .2ex}
    {1ex}

\usepackage{tikz}
\usepackage{tkz-graph}
\usetikzlibrary{decorations.pathmorphing, automata, arrows, shapes,calc}\usetikzlibrary{decorations.pathreplacing}
\usetikzlibrary{automata}
\usetikzlibrary{snakes}
\usetikzlibrary{calc}
\usetikzlibrary{shapes.geometric}
\usetikzlibrary{chains}
\usetikzlibrary{fit}
\usetikzlibrary{trees}
\usetikzlibrary{decorations}
\tikzset{snake/.style={decorate, decoration={snake, segment length=3mm, amplitude=.3mm}}}
\tikzset{double distance=1pt}

\usepackage[bibliography=common]{apxproof} 

\usepackage{booktabs} 

\usepackage{subcaption}

\usepackage{xpatch}
\usepackage{textcase}
\makeatletter
\xpatchcmd{\@sect}{\uppercase}{\MakeTextUppercase}{}{}
\xpatchcmd{\@sect}{\uppercase}{\MakeTextUppercase}{}{}
\makeatother

\newcommand{\inFullVersion}[1]{#1}

\newcommand{\inConfVersion}[1]{}

\newcommand{\forLater}[1]{}

\newcommand{\bag}[1]{\multileft #1 \multiright}

\newcommand{\rpqs}{RPQs\xspace}
\newcommand{\group}[1]{\ensuremath{\textnormal{\textsf{grp}}_{#1}}}

\newcommand{\nc}{\newcommand}

\nc{\OMIT}[1]{}

\nc{\nat}{\mathbb{N}}
\nc{\cP}{\mathcal{P}}
\nc{\cR}{\mathcal{R}}
\nc{\cS}{\mathcal{S}}
\nc{\cA}{\mathcal{A}}
\nc{\cC}{\mathcal{C}}
\nc{\cM}{\mathcal{M}}

\nc{\N}{\mathbb{N}}

\nc{\paths}{\ensuremath{\mathsf{Paths}\xspace}}
\nc{\spaths}{\ensuremath{\mathsf{SPaths}\xspace}}
\nc{\mpaths}{\ensuremath{\mathsf{MPaths}\xspace}}
\nc{\swords}{\ensuremath{\mathsf{SWords}\xspace}}
\nc{\mwords}{\ensuremath{\mathsf{MWords}\xspace}}

\nc{\allmode}{\ensuremath{\mathsf{all}\xspace}}
\nc{\nonemode}{\ensuremath{\mathsf{none}\xspace}}
\nc{\singlemode}{\ensuremath{\mathsf{single}\xspace}}
\nc{\shortestmode}{\ensuremath{\mathsf{shortest}\xspace}}
\nc{\leximode}{\ensuremath{\mathsf{radix}\xspace}}
\nc{\simplemode}{\ensuremath{\mathsf{simple}\xspace}}
\nc{\trailmode}{\ensuremath{\mathsf{trail}\xspace}}

\nc{\src}{\ensuremath{\mathsf{src}}}
\nc{\tgt}{\ensuremath{\mathsf{tgt}}}
\nc{\tab}{\ensuremath{\mathsf{tab}}}
\nc{\pair}{\ensuremath{\mathsf{pair}}}
\nc{\graph}{\ensuremath{\mathsf{graph}}}
\nc{\img}{\ensuremath{\mathsf{Img}}}

\nc{\sfs}{\ensuremath{\mathsf{s}}}
\nc{\sft}{\ensuremath{\mathsf{t}}}
\nc{\sfst}{\ensuremath{\mathsf{st}}}
\nc{\sfp}{\ensuremath{\mathsf{p}}}

\nc{\bL}{\ensuremath{\mathbf{L}\xspace}}
\nc{\bK}{\ensuremath{\mathbf{K}\xspace}}
\nc{\bV}{\ensuremath{\mathbf{V}\xspace}}

\nc{\nid}{\ensuremath{\mathsf{NID}}\xspace}
\nc{\eid}{\ensuremath{\mathsf{EID}}\xspace}
\nc{\pid}{\ensuremath{\mathsf{PID}}\xspace}
\nc{\var}{\ensuremath{\mathsf{Var}}\xspace}
\nc{\pvar}{\ensuremath{\mathsf{PathVar}}\xspace}

\nc{\expr}{\ensuremath{\mathsf{exp}}\xspace}
\nc{\dfa}{\ensuremath{\mathsf{dfa}}\xspace}
\nc{\ufa}{\ensuremath{\mathsf{ufa}}\xspace}

\nc{\be}{\ensuremath{\mathbf{e}\xspace}}
\nc{\bv}{\ensuremath{\mathbf{v}\xspace}}

\nc{\lab}{\ensuremath{\mathsf{lab}}\xspace}
\nc{\true}{\ensuremath{\mathsf{true}}\xspace}
\nc{\false}{\ensuremath{\mathsf{false}}\xspace}

\nc{\pr}{PMR\xspace} \nc{\prs}{PMRs\xspace} \nc{\pmr}{PMR\xspace} \nc{\pmrs}{PMRs\xspace} 

\nc{\grpq}{GRPQ\xspace} \nc{\grpqs}{GRPQs\xspace} 

\nc{\ugrpq}{UGRPQ\xspace} \nc{\ugrpqs}{UGRPQs\xspace} 

\DeclareMathOperator{\lang}{\ensuremath{L}}

\newcommand{\size}[1]{\ensuremath{|#1|}}
\DeclareMathOperator{\bigo}{\ensuremath{\mathcal{O}}}
\DeclareMathOperator{\trim}{trim}

\newcommand{\ptime}{\text{PTIME}\xspace}
\newcommand{\np}{\text{NP}\xspace}
\newcommand{\pspace}{\text{PSPACE}\xspace}

\nc{\fset}{\mathsf{FSet}\xspace}
\nc{\multileft}{\ensuremath{\{\!\!\{}}
\nc{\multiright}{\ensuremath{\}\!\!\}}}

\nc{\prse}{\text{PRSE}\xspace}
\nc{\prme}{\text{PRME}\xspace}
\nc{\prsm}{\text{PRSM}\xspace}
\nc{\prmm}{\text{PRMM}\xspace}

\theoremstyle{plain}
\newtheoremrep{theorem}{Theorem}[section]
\newtheoremrep{lemma}[theorem]{Lemma}
\newtheoremrep{proposition}[theorem]{Proposition}
\newtheoremrep{claim}[theorem]{Claim}
\newtheoremrep{corollary}[theorem]{Corollary}
\newtheoremrep{observation}[theorem]{Observation}
\newtheoremrep{remark}[theorem]{Remark}

\newcommand{\set}{\textsf{set}\xspace}

\definecolor{darkblue}{RGB}{0,56,153}
\definecolor{darkred}{RGB}{153,0,0}
\usetikzlibrary{arrows,positioning,backgrounds}

\def\stoptoken{!}
\newif\ifpathstop
\def\vmedge#1,{\if\stoptoken#1\let\vmnode\relax\else,\textcolor{red}{#1},\fi\vmnode}
\def\vmnode#1,{\if\stoptoken#1\let\vmedge\relax\else\textcolor{blue}{#1}\fi\vmedge}

\definecolor{darkblue}{RGB}{0,56,153}
\definecolor{darkred}{RGB}{153,0,0}
\newcommand{\pathvalue}[1]{\pathpred({\texteltid{#1}})}
\newcommand{\pathpred}{\mathsf{path}}

\newcommand{\eltidfont}{\normalfont\ttfamily}\newcommand{\texteltid}[1]{\text{{\eltidfont #1}}}
\newcommand{\nodeid}[1]{\texteltid{\color{darkred} #1}}
\newcommand{\edgeid}[1]{\texteltid{\color{blue} #1}}

\newcommand{\sem}[1]{\ensuremath{[#1]}_{\mathsf{tab}}}

\newcommand{\semg}[1]{\ensuremath{\mathsf{graph}(#1)}}

\definecolor{BrickRed}{rgb}{0.8, 0.25, 0.33}
\newcommand{\wimColor}{BrickRed}

\newcommand{\wim}[1]{\textcolor{\wimColor}{\textbf{Wim:} #1}}

\newlength\boxwidth
\newlength\questionwidth
\newcommand{\decisionproblem}[4]{
    \setlength\boxwidth{#1-.5cm}{
        \setlength\questionwidth{#1-.5cm}\addtolength\questionwidth{-2.1cm}{
            \begin{center}
                \fbox{\parbox[t]{\boxwidth}{\centerline{#2}
                        \vspace{1mm}
                        \begin{tabular}{l@{\hspace{3mm}}p{\questionwidth}} 
                            Given: & #3\\[1pt]
                            Question: & #4\\
                \end{tabular}}}
            \end{center}
}}}

\newcommand\restr[2]{{\left.\kern-\nulldelimiterspace #1 \vphantom{\big|} \right|_{#2} }}

\AtBeginDocument{\providecommand\BibTeX{{\normalfont B\kern-0.5em{\scshape i\kern-0.25em b}\kern-0.8em\TeX}}}

\setcopyright{acmlicensed}

\begin{document}
\title{Representing Paths in Graph Database Pattern Matching}

\fancyhead{}

\author{Wim Martens}
\affiliation{
  \institution{University of Bayreuth}
  \country{Germany}
}

\author{Matthias Niewerth}
\affiliation{
  \institution{University of Bayreuth}
  \country{Germany}
}

\author{Tina Popp}
\affiliation{
  \institution{University of Bayreuth}
  \country{Germany}
}

\author{Stijn Vansummeren}
\affiliation{
  \institution{Hasselt University}
  \country{Belgium}
}

\author{Domagoj Vrgo\v{c}}
\affiliation{
  \institution{Pontificia Universidad Cat\'olica}
  \country{Chile}
}

 \ccsdesc[500]{Information systems~Query languages for non-relational engines}
 \ccsdesc[500]{Theory of computation~Database query languages (principles)}
 \ccsdesc[300]{Theory of computation~Regular languages}

 \keywords{Graph databases, query languages, query evaluation}

\begin{abstract}
  Modern graph database query languages such as GQL, SQL/PGQ, and their academic
  predecessor G-Core promote
  paths to first-class citizens in the sense that paths that match regular path
  queries can be returned to the user. This brings a number of challenges in
  terms of efficiency, caused by the fact that graphs can have a huge amount of
  paths between a given node pair.

  We introduce the concept of \emph{path multiset representations (PMRs)}, which
  can represent multisets of paths in an exponentially succinct manner. After
  exploring fundamental problems such as minimization and
  equivalence testing of \pmrs, we explore how their use can lead to
  significant time and space savings when executing query plans. We show that,
  from a computational complexity point of view, \pmrs seem especially
  well-suited for representing results of regular path queries and extensions
  thereof involving counting, random sampling, unions, and joins.
\end{abstract}

\maketitle

\section{Introduction}

Graph databases are becoming increasingly popular \cite{SakrBVIAAAABBDV-cacm21}. Indeed,
modern graph query languages such as Neo4j's Cypher~\cite{cypher}, Tigergraph's
GSQL \cite{gsql}, and Oracle's PGQL \cite{pgql} are rapidly gaining adoption in
industry, and there are ongoing ISO standardization efforts for GQL (a
native query language for property graphs) as well as SQL/PGQ (which extends SQL
with capabilities for graph pattern matching on property
graphs)~\cite{GQL-industry}.

At the core of all of these languages lies the problem of evaluating
\emph{regular path queries} (or \emph{RPQs} for short), which have been
studied in database research since the late 1980s, see, e.g., \cite{CruzMW-sigmod87,CalvaneseGLV-pods99,CalvaneseGLV-kr00,Barcelo-pods13,FigueiraGKMNT-kr20,MartensT-tods19,MendelzonW-sicomp95,MartensNT-stacs20,BaganBG-pods13,BarceloLLW-tods12}. In essence, an RPQ consists of a
regular expression $e$. The classical semantics of RPQs in the academic
literature and in, e.g., implementations of SPARQL~\cite{sparql11} is the
following. When we evaluate $e$ over an edge-labeled graph $G$, we return all
node pairs $(x,y)$ such that there exists a path from $x$ to $y$ in $G$ whose
sequence of edge labels forms a word in the language of $e$.  Modern graph query
languages such as GQL, SQL/PGQ, and their academic predecessors such as
G-Core~\cite{gcore}, are adopting a fundamentally different approach
by making paths \emph{first-class citizens}: RPQs no
longer simply return endpoint pairs, but also the matching paths. We illustrate both semantics by means of the following example.

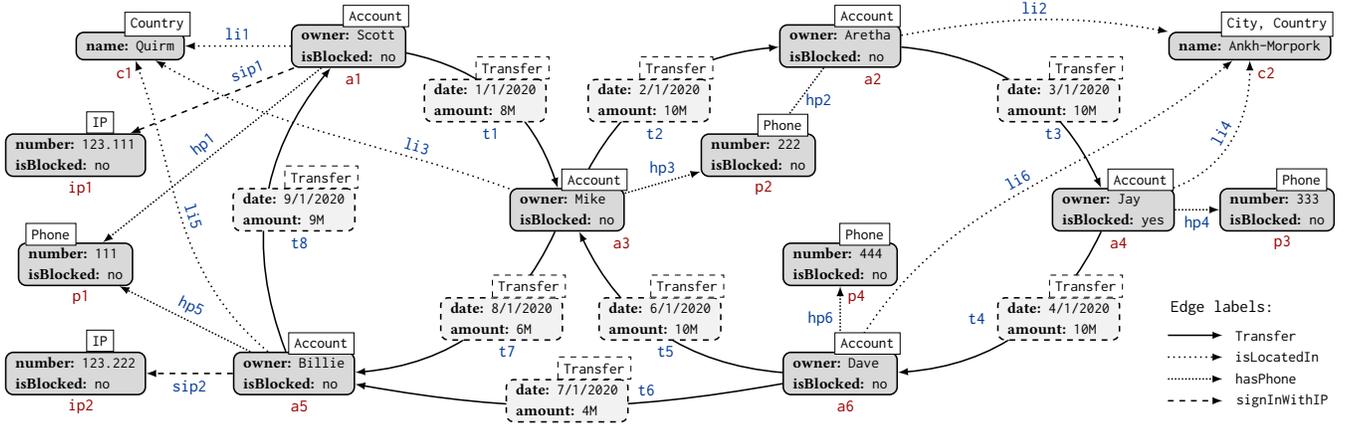
\begin{figure*}
    \resizebox{\linewidth}{!}{
    \begin{tikzpicture}
            \tikzstyle{every state}=[draw,thick,rectangle,rounded corners,fill=white!85!black,minimum size=5mm, text=black, font=\ttfamily, inner sep=0pt]
            \tikzstyle{every node}=[font=\ttfamily]

\node[state] (acc1) at (-4,3) {
          \begin{tabular}{l}
            \small {\bf owner:} Scott\\
            \small {\bf isBlocked:} no\\
          \end{tabular}
        };

        \node[state] (acc2) at (5,3) {
          \begin{tabular}{l}
            \small {\bf owner:} Aretha\\
            \small {\bf isBlocked:} no\\
          \end{tabular}
        };

        \node[state] (acc3) at (0,0) {
          \begin{tabular}{l}
            \small {\bf owner:} Mike\\
            \small {\bf isBlocked:} no\\
          \end{tabular}
        };

        \node[state] (acc4) at (10,0) {
          \begin{tabular}{l}
            \small {\bf owner:} Jay\\
            \small {\bf isBlocked:} yes\\
          \end{tabular}
        };

        \node[state] (acc5) at (-5,-3) {
          \begin{tabular}{l}
            \small {\bf owner:} Billie\\
            \small {\bf isBlocked:} no\\
          \end{tabular}
        };

        \node[state] (acc6) at (5,-3) {
          \begin{tabular}{l}
            \small {\bf owner:} Dave\\
            \small {\bf isBlocked:} no\\
          \end{tabular}
        };

\node[rectangle,draw,fill=white] (acc1label) at ($(acc1)+(0.5,0.55)$) {\ttfamily \small Account};
        \node[rectangle,draw,fill=white] (acc2label) at ($(acc2)+(0.5,0.55)$) {\ttfamily \small Account};
        \node[rectangle,draw,fill=white] (acc3label) at ($(acc3)+(0.5,0.55)$) {\ttfamily \small Account};
        \node[rectangle,draw,fill=white] (acc4label) at ($(acc4)+(0.5,0.55)$) {\ttfamily \small Account};
        \node[rectangle,draw,fill=white] (acc5label) at ($(acc5)+(0.5,0.55)$) {\ttfamily \small Account};
        \node[rectangle,draw,fill=white] (acc6label) at ($(acc6)+(0.5,0.55)$) {\ttfamily \small Account};
      
\node[text=darkred] at ($(acc1)+(.1,-.6)$) {a1};
        \node[text=darkred] at ($(acc2)+(.6,-.6)$) {a2};
        \node[text=darkred] at ($(acc3)+(,-.6)$) {a3};
        \node[text=darkred] at ($(acc4)+(.1,-.6)$) {a4};
        \node[text=darkred] at ($(acc5)+(.1,-.6)$) {a5};
        \node[text=darkred] at ($(acc6)+(.1,-.6)$) {a6};

\node[state,dashed,fill=white!95!black] (t1) at (-1.5,2) {
            \begin{tabular}{l}
                \small {\bf date:} 1/1/2020\\
                \small {\bf amount:} 8M\\
            \end{tabular}
        };
        \node[state,dashed,fill=white!95!black] (t2) at (1.5,2) {
            \begin{tabular}{l}
                \small {\bf date:} 2/1/2020\\
                \small {\bf amount:} 10M\\
            \end{tabular}
        };
        \node[state,dashed,fill=white!95!black] (t3) at (9,2) {
            \begin{tabular}{l}
                \small {\bf date:} 3/1/2020\\
                \small {\bf amount:} 10M\\
            \end{tabular}
        };
        \node[state,dashed,fill=white!95!black] (t4) at (9,-2) {
            \begin{tabular}{l}
                \small {\bf date:} 4/1/2020\\
                \small {\bf amount:} 10M\\
            \end{tabular}
        };
        \node[state,dashed,fill=white!95!black] (t5) at (1.7,-2) {
            \begin{tabular}{l}
                \small {\bf date:} 6/1/2020\\
                \small {\bf amount:} 10M\\
            \end{tabular}
        };
        \node[state,dashed,fill=white!95!black] (t6) at (0,-3.5) {
            \begin{tabular}{l}
                \small {\bf date:} 7/1/2020\\
                \small {\bf amount:} 4M\\
            \end{tabular}
        };
        \node[state,dashed,fill=white!95!black] (t7) at (-1.2,-2) {
            \begin{tabular}{l}
                \small {\bf date:} 8/1/2020\\
                \small {\bf amount:} 6M\\
            \end{tabular}
        };
        \node[state,dashed,fill=white!95!black] (t8) at (-5,0) {
            \begin{tabular}{l}
                \small {\bf date:} 9/1/2020\\
                \small {\bf amount:} 9M\\
            \end{tabular}
        };
      
\node[rectangle,draw,dashed,fill=white] (t1label) at ($(t1)+(0.5,0.6)$) {\ttfamily \small Transfer};
        \node[rectangle,draw,dashed,fill=white] (t2label) at ($(t2)+(0.5,0.6)$) {\ttfamily \small Transfer};
        \node[rectangle,draw,dashed,fill=white] (t3label) at ($(t3)+(0.5,0.6)$) {\ttfamily \small Transfer};
        \node[rectangle,draw,dashed,fill=white] (t4label) at ($(t4)+(0.5,0.6)$) {\ttfamily \small Transfer};
        \node[rectangle,draw,dashed,fill=white] (t5label) at ($(t5)+(0.5,0.6)$) {\ttfamily \small Transfer};
        \node[rectangle,draw,dashed,fill=white] (t6label) at ($(t6)+(0.5,0.6)$) {\ttfamily \small Transfer};
        \node[rectangle,draw,dashed,fill=white] (t7label) at ($(t7)+(0.5,0.6)$) {\ttfamily \small Transfer};
        \node[rectangle,draw,dashed,fill=white] (t8label) at ($(t8)+(0.5,0.6)$) {\ttfamily \small Transfer};

\node[text=darkblue] at ($(t1)+(.1,-.6)$) {t1};
        \node[text=darkblue] at ($(t2)+(.1,-.6)$) {t2};
        \node[text=darkblue] at ($(t3)+(-.1,-.6)$) {t3};
        \node[text=darkblue] at ($(t4)+(-1.5,0)$) {t4};
        \node[text=darkblue] at ($(t5)+(.1,-.6)$) {t5};
        \node[text=darkblue] at ($(t6)+(1.45,.2)$) {t6};
        \node[text=darkblue] at ($(t7)+(.1,-.6)$) {t7};
        \node[text=darkblue] at ($(t8)+(.1,-.6)$) {t8};

\node[state] (p1) at (-9,-1) {
          \begin{tabular}{l}
            \small {\bf number:} 111\\
            \small {\bf isBlocked:} no\\
          \end{tabular}
        };

        \node[state] (p2) at (3.5,1) {
          \begin{tabular}{l}
            \small {\bf number:} 222\\
            \small {\bf isBlocked:} no\\
          \end{tabular}
        };

        \node[state] (p3) at (13,0) {
          \begin{tabular}{l}
            \small {\bf number:} 333\\
            \small {\bf isBlocked:} no\\
          \end{tabular}
        };

        \node[state] (p4) at (5,-1) {
          \begin{tabular}{l}
            \small {\bf number:} 444\\
            \small {\bf isBlocked:} no\\
          \end{tabular}
        };

\node[rectangle,draw,fill=white] (p1label) at ($(p1)+(-0.45,0.55)$) {\ttfamily \small Phone};
      \node[rectangle,draw,fill=white] (p2label) at ($(p2)+(0.45,0.55)$) {\ttfamily \small Phone};
      \node[rectangle,draw,fill=white] (p3label) at ($(p3)+(0.45,0.55)$) {\ttfamily \small Phone};
      \node[rectangle,draw,fill=white] (p4label) at ($(p4)+(0.45,0.55)$) {\ttfamily \small Phone};

\node[text=darkred] at ($(p1)+(.1,-.6)$) {p1};
        \node[text=darkred] at ($(p2)+(.1,-.6)$) {p2};
        \node[text=darkred] at ($(p3)+(.1,-.6)$) {p3};
        \node[text=darkred] at ($(p4)+(.3,-.6)$) {p4};

\node[state] (ip1) at (-9,1) {
          \begin{tabular}{l}
            \small {\bf number:} 123.111\\
            \small {\bf isBlocked:} no\\
          \end{tabular}
        };

        \node[state] (ip2) at (-9,-3) {
          \begin{tabular}{l}
            \small {\bf number:} 123.222\\
            \small {\bf isBlocked:} no\\
          \end{tabular}
        };

\node[rectangle,draw,fill=white] (ip1label) at ($(ip1)+(0.45,0.6)$) {\ttfamily \small IP};
      \node[rectangle,draw,fill=white] (ip2label) at ($(ip2)+(0.45,0.6)$) {\ttfamily \small IP};
      
\node[text=darkred] at ($(ip1)+(.1,-.6)$) {ip1};
        \node[text=darkred] at ($(ip2)+(.1,-.6)$) {ip2};

\node[state] (c1) at (-8,3) {
          \begin{tabular}{l}
            \small {\bf name:} Quirm\\
          \end{tabular}
        };
      
        \node[state] (c2) at (12.5,3) {
          \begin{tabular}{l}
            \small {\bf name:} Ankh-Morpork\\
          \end{tabular}
        };

\node[rectangle,draw,fill=white] (c1label) at ($(c1)+(0.5,0.4)$) {\ttfamily \small Country};
        \node[rectangle,draw,fill=white] (c2label) at ($(c2)+(0.5,0.4)$) {\ttfamily \small City, Country};

\node[text=darkred] at ($(c1)+(-.1,-.5)$) {c1};
        \node[text=darkred] at ($(c2)+(.3,-.5)$) {c2};

\begin{pgfonlayer}{background}        
\path[-latex,thick]
        (acc1) edge[bend left] (acc3)
        (acc3) edge[bend left] (acc2)
        (acc2) edge[bend left] (acc4)
        (acc4) edge[bend left] (acc6)
        (acc6) edge[out=190,in=-10] (acc5)
        (acc6) edge[bend left] (acc3)
        (acc3) edge[bend left] (acc5)
        (acc5) edge[bend left] (acc1)
        ;
          
\path[-latex,thick,densely dotted]
        (acc1) edge node [above,sloped,text=darkblue] {hp1} (p1)
        (acc2) edge node [right,text=darkblue] {hp2} (p2)
        (acc3) edge node [above,text=darkblue] {hp3} (p2)
        (acc4) edge node [below,text=darkblue] {hp4} (p3)
        (acc5) edge node [above,sloped,text=darkblue] {hp5} (p1)
        (acc6) edge node [left,text=darkblue] {hp6} (p4)
        ;
        
\path[-latex,thick,dotted] 
        (acc1) edge [out=180,in=0] node [above,sloped,text=darkblue] {li1} (c1)
        (acc2) edge [out=10,in=170] node [above,sloped,text=darkblue] {li2} (c2)
        (acc3) edge [out=160,in=-30] node [above,sloped,near start,text=darkblue] {li3} (c1)
        (acc4) edge [out=20,in=270] node [above,sloped,text=darkblue] {li4} (c2)
        (acc5) edge [out=140,in=290] node [above,sloped,text=darkblue] {li5} (c1)
        (acc6) edge [out=60,in=220] node [above,sloped,text=darkblue] {li6} (c2)
        ;

\path[-latex,thick,dashed]
        (acc1) edge node [above,sloped,near start,text=darkblue] {sip1} (ip1)
        (acc5) edge node [below,text=darkblue] {sip2} (ip2)
        ;
    \end{pgfonlayer}

\node at (12,-1.8) {Edge labels:};
    
    \path[-latex,thick] (11,-2.3) edge (12,-2.3);
    \node at (12.8,-2.3) {\small{\texttt{Transfer}}};

    \path[-latex,thick, dotted] (11,-2.7) edge (12,-2.7);
    \node at (13,-2.7) {\small{\texttt{isLocatedIn}}};

    \path[-latex,thick,densely dotted] (11,-3.1) edge (12,-3.1);
\node at (12.8,-3.1) {\small{\texttt{hasPhone}}};

    \path[-latex,thick,dashed] (11,-3.5) edge (12,-3.5);
    \node at (13.1,-3.5) {\small{\texttt{signInWithIP}}};
    \end{tikzpicture}
    }
    \caption{A property graph with information on bank accounts, their location,
      and financial transations, based on \cite{GQL-industry}.\label{fig:propertygraph}}
  \end{figure*}

\begin{example}\label{ex:intro}
  We adopt the property graph of Figure~\ref{fig:propertygraph}
as our running example. The
  graph has \emph{node identifiers} (\nodeid{a1}, \dots, \nodeid{a6},
  \nodeid{c1}, \nodeid{c2}, \nodeid{p1},\dots,\nodeid{p4}, \nodeid{ip1},
  \nodeid{ip2}) in red and \emph{edge identifiers} (\edgeid{t1}, \dots,
  \edgeid{t8}, \edgeid{li1}, \dots, \edgeid{li6}, \edgeid{hp1}, \dots,
  \edgeid{hp6}) in blue. Nodes and edges can carry \emph{labels} (such as
  \textsf{Account}, \textsf{Transfer}, and \textsf{isLocatedIn}) and
  \emph{property-value} pairs (such as (\textsf{owner}, \textsf{Mike}) and
  (\textsf{date}, \textsf{1/1/2020})). We depict labels and
  property/value pairs for nodes in solid boxes, whereas for edges, these are in
  dashed boxes (or in the legend on the bottom right).

  Consider the RPQ consisting of the regular expression
  $e = \textsf{Transfer}^{\textsf{+}}.$
  When evaluated under the classical
  semantics on the graph in Figure~\ref{fig:propertygraph}, this RPQ returns all
  node pairs $(x,y)$ such that there is a path of length at least one from $x$
  to $y$ in which every edge carries the label \textsf{Transfer}. Examples of
  such node pairs are $(\nodeid{a1},\nodeid{a3})$ (which have a direct
  $\textsf{Transfer}$ link) but also $(\nodeid{a1},\nodeid{a2})$ (connected by a
  path of length $2$) and $(\nodeid{a1},\nodeid{a4})$ (connected by a path of
  length $3$).  When evaluated under the new semantics, however, this RPQ
  would also return the matching paths in addition to the endpoint pairs, and
  include answers like
  \begin{align*}
&(\nodeid{a1},\nodeid{a3},\ \pathvalue{\nodeid{a1},\edgeid{t1},\nodeid{a3}}),\\ 
&(\nodeid{a1},\nodeid{a2},\ \pathvalue{\nodeid{a1},\edgeid{t1},\nodeid{a3},\edgeid{t2},\nodeid{a2}}),\\ 
&(\nodeid{a1},\nodeid{a4},\ \pathvalue{\nodeid{a1},\edgeid{t1},\nodeid{a3},\edgeid{t2},\nodeid{a2},\edgeid{t3},\nodeid{a4}}),
\end{align*}
  where we used
$\pathvalue{\nodeid{a1},\edgeid{t1},\nodeid{a3},\edgeid{t2},\nodeid{a2}}$
  to denote the shortest
  path from Scott ($\nodeid{a1}$) to Aretha ($\nodeid{a2}$) in
  Figure~\ref{fig:propertygraph}. Under this semantics, RPQs hence return
  triples $(x,y,p)$ where $x$ and $y$ are nodes and $p$ is a path that connects them.
\qed
\end{example}

Making paths a first-class citizen in modern graph query languages is not a
straightforward task. Fundamentally, a key problem that systems are facing is how to best \emph{represent} results of queries and subqueries that feature paths. The main issue is dealing with the sheer number of results that path queries can produce, and how to present these to the user.

A common approach, proposed by the GQL standard~\cite{GQL-industry}, by
SQL/PGQ~\cite{GQL-industry}, and already supported by multiple
engines~\cite{FrancisGGLLMPRS-sigmod18,neo,tigergraph,Stardog,MillenniumDB}, is to return the results of a
path query as a relational table. For instance, in Example \ref{ex:intro}, this
table contains triples $(x,y,p)$, where $x$ and $y$ are nodes, while $p$ is a
$\textsf{Transfer}$-labeled path connecting them. We show a portion of this
table (replacing node/edge IDs with their content for readability) in
Table~\ref{tab:tabularoutput-a}. However, the number of results, when represented
in such a way, can quickly become prohibitively large, or even infinite. To
illustrate this, notice that the graph in Figure~\ref{fig:propertygraph} has
several $\textsf{Transfer}$-labeled cycles. This, in turn, implies that there
is an infinite number of triples $(x,y,p)$, where $x$ and $y$ are nodes
connected by a $\textsf{Transfer}$-labeled path. For instance, there are
infinitely many paths between $\nodeid{a1}$ and $\nodeid{a3}$, of lengths 1, 5,
9, etc. To ensure that queries have finite answers, GQL and the existing query
engines restrict the paths that are allowed. Common types of paths considered
are: \textsf{TRAIL} (no repeated edge), \textsf{SIMPLE} (no repeated node), and
\textsf{SHORTEST}~\cite{GQL-industry}.

\begin{table*}
  \begin{subtable}[h]{0.45\textwidth}
    \small
    \begin{tabular}{llp{5.5cm}}
      \toprule
      $x$ & $y$ & $p$\\
      \midrule
      \texttt{Mike} & \texttt{Billie} & \texttt{Mike -[Transfer]-> Billie}\\
      \texttt{Billie} & \texttt{Scott} & \texttt{Billie -[Transfer]-> Scott}\\
      \texttt{Scott} & \texttt{Mike} & \texttt{Scott -[Transfer]-> Mike}\\
      \texttt{Mike} & \texttt{Aretha} & \texttt{Mike -[Transfer]-> Aretha}\\
\texttt{[...]}\\
      \texttt{Mike} & \texttt{Aretha} & \texttt{Mike -[Transfer]-> Billie -[Transfer]->} \\
          & & \texttt{\ Scott -[Transfer]-> Mike -[Transfer]->}\\
      & & \texttt{\ Aretha }\\
      \texttt{Mike} & \texttt{Billie} & \texttt{Mike -[Transfer]-> Billie -[Transfer]->}\\
      & & \texttt{\ Scott -[Transfer]-> Mike -[Transfer]->  }\\
          & & \texttt{\ Aretha -[Transfer]-> Jay -[Transfer]-> }\\
      & & \texttt{\ Dave -[Transfer]-> Billie }\\
      \texttt{[...]}\\
    \bottomrule
    \end{tabular}
    \caption{Tabular representation of
      \texttt{Transfer}-trails in Figure~\ref{fig:propertygraph}.\label{tab:tabularoutput-a}}
  \end{subtable}
  \quad
  \begin{subtable}[h]{0.52\textwidth}
    \small \centering
  \begin{tabular}{llp{5.5cm}}
      \toprule
      $x$ & $y$ & $p$\\
      \midrule
      \texttt{Mike} & \texttt{Billie} & \texttt{Mike -[Transfer]-> Billie}, \\
      & & \texttt{Mike -[Transfer]-> Billie -[Transfer]->}\\
      & & \texttt{\ Scott -[Transfer]-> Mike -[Transfer]->  }\\
          & & \texttt{\ Aretha -[Transfer]-> Jay -[Transfer]-> }\\
          & & \texttt{\ Dave -[Transfer]-> Billie},\\
      & & \texttt{[...]}\\
      \texttt{Mike} & \texttt{Aretha} & \texttt{Mike -[Transfer]-> Aretha},\\
      & & \texttt{Mike -[Transfer]-> Billie -[Transfer]->} \\
          & & \texttt{\ Scott -[Transfer]-> Mike -[Transfer]->}\\
      & & \texttt{\ Aretha},\\
      & & \texttt{[...]}\\
\texttt{[...]}\\
      \phantom{\texttt{[...]}}\\
      \bottomrule
  \end{tabular}
  \caption{Pairwise grouped tabular representation of
    \texttt{Transfer}-trails in Figure~\ref{fig:propertygraph}.\label{tab:tabularoutput-b}}
  \end{subtable}
  \caption{\label{tab:tabularoutput}Tabular representation of trails.}
\end{table*}

\begin{figure}[t] \centering
		\begin{tikzpicture}[auto,>=latex,->]
		\def\y{.3}	
		
			\node [](q0) at (0,0*\y) [draw,rectangle,rounded corners] {$x$}; 
			\node [](q1) at (1,1*\y)[draw,rectangle,rounded corners] {$u_1$}; 
			\node [](q2) at (1,-1*\y) [draw,rectangle,rounded corners] {$v_1$}; 	
			\node [] (q3) at (2,0*\y) [draw,rectangle,rounded corners] {\phantom{$v_1$}};	
			\node [](q4) at (3,1*\y) [draw,rectangle,rounded corners] {$u_2$}; 
			\node [](q5) at (3,-1*\y) [draw,rectangle,rounded corners] {$v_2$}; 
			\node [](q6) at (4,0*\y) [draw,rectangle,rounded corners] {\phantom{$v_2$}}; 
			
			\node [](q7) at (5,0*\y) [] {$\cdots$}; 
			
			\node [](q7) at (6,0*\y) [draw,rectangle,rounded corners] {\phantom{$v_n$}}; 
			\node [](q8) at (7,1*\y)[draw,rectangle,rounded corners] {$u_n$}; 
			\node [](q9) at (7,-1*\y) [draw,rectangle,rounded corners] {$v_n$}; 
			\node [](qn) at (8,0*\y) [draw,rectangle,rounded corners] {$y$};

			\path (q0) edge   node {$a$}  (q1);
			\path (q0) edge  [swap] node {$a$} (q2);
			\path (q1) edge   node  {$a$} (q3);
			\path (q2) edge  [swap] node  {$a$} (q3);
			
			\path (q3) edge   node {$a$}  (q4);
			\path (q3) edge  [swap] node {$a$} (q5);
			\path (q4) edge   node  {$a$} (q6);
			\path (q5) edge  [swap] node  {$a$} (q6);
			
			\path (q7) edge   node {$a$}  (q8);
			\path (q7) edge  [swap] node {$a$} (q9);
			\path (q8) edge   node  {$a$} (qn);
			\path (q9) edge  [swap] node  {$a$} (qn);
		\end{tikzpicture}
	\caption{A graph with $2^n$ shortest paths from $x$ to~$y$. }
	\label{fig:2n-shortest-in-graph}
\end{figure}
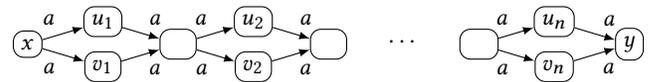

While these evaluation modes do fix the infinity issue, they can still result in
prohibitively large outputs. To illustrate this, consider now the graph in
Figure~\ref{fig:2n-shortest-in-graph}, which has $3n + 1$ nodes and $4n$ edges.
If we were to output all the shortest paths between $x$ and $y$ in
Figure~\ref{fig:2n-shortest-in-graph}, there are $2^n$ of these. Notice that
these paths are also both trails and simple paths. Therefore, a relational table
representation of this output, such as the one in
Table~\ref{tab:tabularoutput-a}, would require to ``materialize'' all $2^n$
paths. For this reason, it seems desirable to adopt a different data structure
that can represent sets of triples $(x,y,p)$ as succinctly as possible,
preferably in less than $2^n$ space, while still allowing to generate the
relational table representation from
them. 

Since the relational table representation can overwhelm the
user, some query engines such as Neo4J~\cite{cypher} present query
results by means of so-called \emph{graph projections}. Intuitively speaking,
the graph projection takes the table representation and displays the subgraph of
the original graph consisting only of the nodes and edges mentioned in the
table. For instance, the graph projection of the query that asks for all paths
from node $x$ to node $y$ in Figure~\ref{fig:2n-shortest-in-graph} simply yields
the graph of Figure~\ref{fig:2n-shortest-in-graph} itself. Although graph
projections can indeed provide users with a compact visualization of the query
result, current system still use the table itself (which can be exponentially
larger than the projection) to compute the graph projection from.
Furthermore, graph projections are not lossless --- they are just a subgraph of
the input, and as such they lose the information about which paths were to be
returned. In Figure~\ref{fig:2n-shortest-in-graph}, the graph projection of the
$2^n$ paths is the same as the projection of the two paths where one goes
through $u_1, u_2, \ldots, u_n$ and the other through $v_1,v_2,\ldots,v_n$.

In this paper we present a conceptual tool for representing (multi)sets of paths
in a compact way, both when this set is infinite, or exponentially large, and
show how this representation can be used to represent intermediate results when
a path query is part of a larger graph query.

\subsection*{Our Contribution}
We introduce the concept of \emph{path multiset representations (\prs)} and
present evidence that query processing based on \prs can make query evaluation
drastically more efficient. In a nutshell, \prs aim to combine the best of the
relational table representation and graph projections while avoiding their
disadvantages. That is, they provide a compact and lossless representation to an
exponential (or even infinite) number of results, similarly as a graph
projection, while at the same time allowing to identify individual paths in the
output, as the tabular representation does. Intuitively, a \pr over a graph $G$
is itself a graph $R$, together with
\begin{itemize}
\item a homomorphism $\gamma$ from $R$ to $G$, and
\item a set of ``start nodes'' $S$ and ``target nodes'' $T$.
\end{itemize}
The idea is that $R$ provides a succinct structure to represent paths between
groups of nodes in $G$.

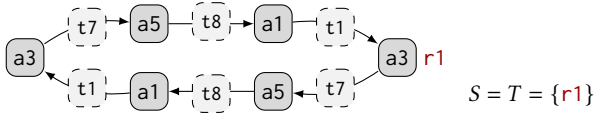
\begin{figure}
\begin{tikzpicture}
			\tikzstyle{every state}=[draw,rectangle,rounded corners,fill=white!85!black,minimum size=5mm, text=black, font=\ttfamily, inner sep=0pt]
			\tikzstyle{every node}=[font=\ttfamily]
			\def\x{.55}
			\def\y{.15}
			
\node[state] (acc1) at (\x*-4,\y*3) {
				a1
			};
			
			\node[state] (acc3) at (\x*-1,\y*0) {
				a3
			};
			
			\node[state] (acc5) at (\x*-4,\y*-3) {
				a5
			};
			
			\node[state] (acc11) at (\x*-7,\y*-3) {
				a1
			};
			
			\node[state] (acc13) at (\x*-10,\y*0) {
				a3
			};
			
			\node[state] (acc15) at (\x*-7,\y*3) {
				a5
			};
			
\node[text=darkred] at ($(acc3)+(.5,0)$) {r1};

\node[state,dashed,fill=white!95!black] (t1) at (\x*-2.5,\y*2.7) {
			\small 	t1
			};
			\node[state,dashed,fill=white!95!black] (t7) at (\x*-2.5,\y*-2.7) {
			\small 	t7
			};
			\node[state,dashed,fill=white!95!black] (t8) at (\x*-5.5,\y*3.2) {
			\small 	t8
			};
			
			\node[state,dashed,fill=white!95!black] (t11) at (\x*-8.5,\y*-2.7) {
			\small 	t1
			};
			\node[state,dashed,fill=white!95!black] (t17) at (\x*-8.5,\y*2.7) {
			\small 	t7
			};
			\node[state,dashed,fill=white!95!black] (t18) at (\x*-5.5,\y*-3.2) {
			\small 	t8
			};

			\begin{pgfonlayer}{background}        
\path[-latex]
				(acc1) edge[bend left=20] (acc3)
				(acc3) edge[bend left=20] (acc5)
				(acc5) edge[bend left=0] (acc11)
				(acc11) edge[bend left=20] (acc13)
				(acc13) edge[bend left=20] (acc15)
				(acc15) edge[bend left=0] (acc1)
				;
			\end{pgfonlayer}
		\end{tikzpicture}
  \begin{tikzpicture}
    \node {$S = T = \{\textcolor{darkred}{\texttt{r1}}\}$};
  \end{tikzpicture}
	\caption{Path representation over the graph database of Figure~\ref{fig:propertygraph}, representing  all cycles of even length from Mike to Mike
    where the transferred amounts are strictly less than 10M. 
    We depicted the value $\gamma(u)$ inside every node
    $u$. \label{fig:mike-mike-even}}
\end{figure}

To illustrate, Figure~\ref{fig:mike-mike-even} shows a \pmr $R$
over the graph $G$ of Figure~\ref{fig:propertygraph}. It uses a single start
node, \nodeid{r1}, which is also the single target node, and represents all
cycles of even length from Mike to Mike where the transferred amounts are less
than 10M.  Intuitively, the homomorphism $\gamma$ associates each node in $R$ to
a node in $G$ --- for each node $u$ of $R$, we depicted the value of $\gamma(u)$
inside the node $u$ in Figure~\ref{fig:mike-mike-even}. Notice that $\gamma$ can
associate multiple nodes in $R$ to the same node in $G$. In particular, the leftmost and
rightmost node in $R$ are both mapped to \nodeid{a3}. This symbolizes the fact
that one needs to traverse the cycle
\nodeid{a3}--\nodeid{a5}--\nodeid{a1}--\nodeid{a3} twice to obtain even
length.

A \pmr $R$ ``represents'' a (possibly infinite) number of paths in
$G$. These paths are the images of the paths in $R$ from some node in $S$ to some node
in $T$ under the mapping $\gamma$. As such, in
Figure~\ref{fig:mike-mike-even}, the paths from $S = \{\nodeid{r1}\}$ to $T = \{\nodeid{r1}\}$ are
cycles of length 0, 6, 12, etc.\ in $R$, which correspond (through $\gamma$) to
cycles of the same lengths in $G$. In this case, the number of paths represented
by $R$ is infinite.

In this paper, we formally introduce this novel concept of path multiset
representations (\prs) and explore their use for pattern matching in
modern graph query languages. We obtain a large number of results that show
significant potential:
\begin{enumerate}
\item \prs represent sets and multisets of paths accurately and exponentially
  more succinctly than current state-of-the-art systems (and what is described
  in the current GQL standard).
\item Testing whether two \prs represent the same multiset of paths can be done
  efficiently.
\item \prs for outputs of \emph{regular path queries}, the basic building block
  of modern graph pattern matching languages, can be computed in linear time
  combined complexity, which strongly contrasts with the current exponential
  algorithms.
\item \prs for RPQ outputs can be efficiently converted to graph projections, and to the tabular representation of the output.
\item \prs fit into the framework of larger queries, extending RPQs with unions
  and conjunctions, allowing exponentially more efficient computations in query plans
  due to their exponentially succinct represenation.
\end{enumerate}
We note that this paper is conceptual and aimed at providing theoretical
foundations. In doing so, our hope is to inspire the community to build
practical evaluation methods around the concept of \prs, and empirically
validate them. Nevertheless, the presented results already show that significant
(i.e., exponential) speed-ups in query evaluation methods are theoretically
possible. Furthermore, they show that it is possible to represent infinitely
many paths in query evaluation plans using a finite object, which opens up
further possibilities for the future design of graph query languages.

The structure of the paper is as follows.  We provide mathematical background in Section~\ref{sec:preliminaries}. In Section~\ref{sec:path-mult-repr} we formally define Path Multiset Representations and study their basic properties.
In Section~\ref{sec:grpqs} we introduce (unions of) Generalized Regular Path Queries (\grpqs) as a formal model of classical Regular Path Queries that also return paths. In Section~\ref{sec:answering-grpqs} we show how to evaluate (U)\grpqs using Path Multiset Representations. We deal with conjunctions of \grpqs in Section~\ref{sec:gcrpqs}. We discuss related work in Section~\ref{sec:related-work} and conclude in Section~\ref{sec:conclusions}.
\inConfVersion{Because of space limitations, some details and proofs are omitted. A full version of this paper, whose Appendix contains those items, is available online~\cite{paths-full-paper-version}.}
\inFullVersion{Because of space limitations, some formal statements and proofs are deferred to the Appendix.}

\section{Preliminaries}
\label{sec:preliminaries}

\noindent \textbf{Background.}
For a natural number $n$, we denote the set $\{1,\ldots,n\}$ by $[n]$. A
\emph{multiset} $M$ is a function from a set $S$ to $\nat \cup \{\infty\}$.
We denote multisets
using double braces, e.g., in the multiset $M = \multileft a, a, b \multiright$, we
have that $M(a) = 2$ and $M(b) = 1$. We do not distinguish between sets and multisets where all elements have multiplicity one: i.e., we equate $\multileft a, b \multiright = \{ a, b\}$. For a multiset $M$ we denote by $\set(M)$ the set obtained from $M$ by forgetting multiplicities. For
instance, $\set(\multileft a,a,b\multiright) = \{a,b\} = \multileft a,b \multiright$. 

\smallskip
\noindent \textbf{Graph databases.}
We assume that we have infinite \emph{disjoint} sets $\nid$ of \emph{node
  identifiers}, $\eid$ of \emph{edge identifiers}, and $\bL$ of
\emph{labels}.

Because our focus in this paper will be on how \emph{paths as first-class
  citizens} interact with  regular path queries on graph databases,
we adopt a  formal data model that is a simplified version of property graphs in which
property graph features that are non-essential to our discussion, such as node
labels and property-value records, are omitted. We stress that this is only for
ease of exposition: all of these features can be added to our approach without
influencing our results.
Formally, our data model is an edge-labeled directed multigraph, defined as follows.\begin{definition}\label{def:graphdb}
  A \emph{graph database} is a tuple $G = (N,E,\eta,\lambda)$, where
  \begin{enumerate}[(1)]
\item $N\subseteq \nid$ is a finite set of node identifiers  and $E\subseteq \eid$ is a finite set of edge
    identifiers;
  \item $\eta \colon E \to (N \times N)$ is a total function, called the \emph{incidence mapping}, that associates each edge to the nodes it connects;
  \item $\lambda \colon E \to \bL$ is a total function, called the \emph{labeling function}, that associates a label to each edge. \end{enumerate}
\end{definition}
In what follows, if $G$ is a graph then we will write $N_G$ for the set of $G$'s nodes, and similarly write $E_G$, $\eta_G$, $\lambda_G$ for the set of $G$'s edges, incidence mapping, and labeling function. We may omit  subscripts if $G$ is clear from the context.

An \emph{unlabeled graph} is a triple $(N, E, \eta)$ defined exactly as a graph database, except that the labeling function $\lambda$ is missing.

\smallskip\noindent \textbf{Paths.}
A \emph{path} in a graph database $G$ is a sequence 
\[\rho = v_0 e_1 v_1 e_2 v_2 \cdots e_n v_n\]
with $n \geq 0$, $e_i \in E$, and $\eta(e_i) = (v_{i-1},v_i)$ for every $i \in
[n]$. For readability, we sometimes write $\pathpred(\rho)$ instead of simply $\rho$ to stress that we are talking about a path.
For example
$\pathvalue{\nodeid{a1},\edgeid{t1},\nodeid{a3},\edgeid{t2},\nodeid{a2}}$
is the path of length two from Scott to Aretha in
Figure~\ref{fig:propertygraph}. We use $\paths(G)$ to denote the set of paths in $G$.

If $\rho$ is a path in $G$ and $\lambda$ is $G$'s labeling function, then we write $\lambda(\rho)$ for the sequence of edge labels $\lambda(\rho) = \lambda(e_1) \cdots \lambda(e_n)$ occurring on the edges of $\rho$. We write $\src(\rho)$ for the node $v_0$ at which $\rho$ starts, and $\tgt(\rho)$ for the node $v_n$ at which it ends.
Given two sets of nodes $S$ and $T$, we say that $\rho$ is a path \emph{from $S$
  to $T$} if $\src(\rho) = v_0 \in S$ and $\tgt(\rho) = v_n \in T$.

A \emph{path multiset} over $G$ (or PM over $G$ for short) is a multiset of
paths, all in the same graph $G$. We will often simply speak about path multisets without referring to the graph that they are drawn from, which will be implicit from the context.

 \section{Path Multiset Representations}
\label{sec:path-mult-repr}
\begin{toappendix}
  \label{app:path-mult-repr}
\end{toappendix}

To the best of our knowledge, intermediate or final results of queries in
current graph database query languages such as
Cypher~\cite{cypher}, G-Core~\cite{gcore}, and SQL-PGQ~\cite{GQL-industry} are
always represented as tables in which each path is listed explicitly, 
essentially as in
Table~\ref{tab:tabularoutput-a}.
Our focus is on representing the
\emph{path multisets} involved in query answers in a drastically more succinct
manner.
\begin{example}\label{ex:succinct}
	Consider the set of all paths from $x$ to $y$ in
  Figure~\ref{fig:2n-shortest-in-graph}. Since there are $2^n$ such paths,
  representing them as in Table~\ref{tab:tabularoutput} would take $2^n$ rows.
  Instead, we next propose to represent this set of paths by means of the graph
  in Figure~\ref{fig:2n-shortest-in-graph} itself, together with the set $\{x\}$
  of source nodes and $\{y\}$ of target nodes. This representation has size $O(n)$ instead of $\Omega(2^n)$.
\end{example}

More precisely, we propose to use \emph{path multiset
  representations of $G$}, which we define next.

\begin{definition}
  A \emph{path multiset representation (\pmr) over graph $G$} is a tuple $R = (N,E,\eta,\gamma,S,T)$, where
  \begin{enumerate}
  \item $(N, E, \eta)$ is an unlabeled graph;
  \item $\gamma \colon (N \cup E) \to (N_G \cup E_G)$ is a (total) homomorphism, i.e. a
    function that maps nodes in $R$ to nodes in $G$ and edges in $R$ to edges in $G$ such that, if an edge
$e \in E$  connects $v_1$ to $v_2$ in $R$, then $\gamma(e)$ connects $\gamma(v_1)$ to
$\gamma(v_2)$ in $G$; and
  \item $S,T \subseteq N$ are sets of \emph{source} and \emph{target} nodes, respectively.
  \end{enumerate}
  If $R$ is a \pr, then we sometimes write $N_R$ for its set of nodes, and similarly $E_R$, $\eta_R$, $\gamma_R$, $S_R$, and $T_R$ for the other components. If $R_1$ and $R_2$ are \prs over the same graph $G$ whose nodes and edges are disjoint, then we write $R_1 \sqcup R_2$ for \pr over $G$ obtained by taking the disjoint union of $R_1$ and $R_2$ (defined in the obvious way by taking the union of each component).
\end{definition}
If $R$ is a \pr of $G$, we say that node $v \in N$ \emph{represents} the node
$\gamma(v)$ in $G$. Furthermore, each path
$$\rho = v_0 e_1 v_1 e_2 v_2 \cdots e_n v_n$$
from $S$ to $T$ in $R$ \emph{represents} a path in $G$, namely the path
$$\gamma(\rho) := \gamma(v_0) \gamma(e_1) \gamma(v_1) \gamma(e_2) \gamma(v_2) \cdots \gamma(e_n)
\gamma(v_n)\;.$$

We define $\spaths(R)$ and $\mpaths(R)$ to be the set, resp. multiset, of paths
represented by $R$, that is,
\begin{align*}
  \spaths(R) & := \{\gamma(\rho) \mid \rho \text{ is a path from } S
               \text{ to } T \text{ in } R\}, \\
  \mpaths(R) &  := \multileft \gamma(\rho) \mid \rho \text{ is a path from } S \text{ to } T \text{ in } R\multiright.
\end{align*}
 A \pr $R$
\emph{represents} a multiset $M$ of paths if $M = \mpaths(R)$. It represents a
set of paths $P = \{\rho_1,\rho_2,\ldots\}$ if $P = \spaths(R)$. Notice that, if
$M = \mpaths(R)$, then we always have that $\set(M) = \spaths(R)$. In other
words, if a \pr represents a multiset of paths, it also
represents the corresponding set of paths.

A \pr $R$ is \emph{trim} if every node in $N_R$ is on some path from some node
in $S$ to some node in $T$. Unless mentioned otherwise, we always assume that
\prs are trim.

\subsection{Examples of \pmrs}
If $\gamma$ is the identity function, then a \pr is structurally a
subgraph of $G$. This is already useful, as we illustrated in
Example~\ref{ex:succinct}. By choosing a different $\gamma$, however, we can
incorporate \emph{state information}, which is necessary for evaluating regular
path queries (Example~\ref{ex:state-information}), and \emph{multiplicities} of
paths (Example~\ref{ex:multisets}).

\begin{example}[State information]\label{ex:state-information}
	Figure~\ref{fig:mike-mike-even} shows a \pr $R$ for all cycles
  of even length from Mike to Mike, and where all transferred amounts are strictly less
  than 10M. (We omitted node and edge IDs that are irrelevant.) The ``even
  length'' condition can be encoded in $R$, since $\gamma$ can map different nodes
  in $R$ to the same node in $G$. 
\end{example}

Example~\ref{ex:state-information} illustrates another interesting property of
\prs: they can represent an infinite number of paths in a finite
manner. Indeed, the set of cycles of even length from Mike to Mike in
Example~\ref{ex:state-information} is infinite. We have cycles of length 6, 12,
18, etc.

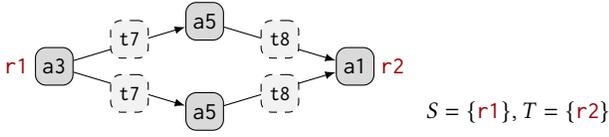
\begin{figure}[t] \centering
		\begin{tikzpicture}
	\tikzstyle{every state}=[draw,rectangle,rounded corners,fill=white!85!black,minimum size=5mm, text=black, font=\ttfamily, inner sep=0pt]
	\tikzstyle{every node}=[font=\ttfamily]
	\def\x{2}
	\def\y{.6}
	
\node[state] (acc1) at (\x*0,\y*0) {a3};	
	\node[state] (acc2) at (\x*1,\y*1) {a5};
	\node[state] (acc3) at (\x*1,\y*-1) {a5};
	\node[state] (acc4) at (\x*2,\y*0) {a1};
	
\node[text=darkred] at ($(acc1)+(-.5,0)$) {r1};
\node[text=darkred] at ($(acc4)+(.5,0)$) {r2};
	
\node[state,dashed,fill=white!95!black] (t1) at (\x*.5,\y*.6) {\small t7};
	\node[state,dashed,fill=white!95!black] (t2) at (\x*.5,\y*-.6) {\small t7};
	\node[state,dashed,fill=white!95!black] (t3) at (\x*1.5,\y*.6) {\small t8};
	\node[state,dashed,fill=white!95!black] (t4) at (\x*1.5,\y*-.6) {\small t8};

	\begin{pgfonlayer}{background}        
\path[-latex]
		(acc1) edge (acc2)
		(acc1) edge (acc3)
		(acc2) edge (acc4)
		(acc3) edge (acc4)
		;
	\end{pgfonlayer}
	
\end{tikzpicture}
\begin{tikzpicture}
	\node {$S = \{\textcolor{darkred}{\texttt{r1}}\}$, $T = \{\textcolor{darkred}{\texttt{r2}}\}$};
\end{tikzpicture}
\caption{Path representation over the graph database of
  Figure~\ref{fig:propertygraph}, representing the path of length two from Mike
  to Scott twice.  We depicted the value $\gamma(u)$ inside every node $u$.  }
	\label{fig:multisets}
\end{figure}

\begin{example}[Multisets]\label{ex:multisets}
	The \pr $R$ in Figure~\ref{fig:multisets} represents the path
  of length two from Mike to Scott twice. We have that $\spaths(R)
  =\{\pathvalue{\nodeid{a3},\edgeid{t7},\nodeid{a5},\edgeid{t8},\nodeid{a1}}\}$
  and $\mpaths(R) =\multileft
  \textsf{path}(\nodeid{a3},\edgeid{t7},\nodeid{a5},\edgeid{t8}, \allowbreak \nodeid{a1}),\pathvalue{\nodeid{a3},\edgeid{t7},\nodeid{a5},\edgeid{t8},\nodeid{a1}} \multiright$.
\end{example}

To represent results of queries (or intermediate results in query plans), our
aim is to work with \pmrs $R$ such that $\mpaths(R)$ corresponds
to the multiset semantics of the query.
We discuss how this is done for regular path queries in Section~\ref{sec:grpqs}
and for conjunctive regular path queries in Section~\ref{sec:gcrpqs}.

\subsection{Basic Properties of \pmrs}
\label{sec:basic-repr}

First, we make some easy but important observations about the kinds of path
multisets that can be represented by \prs. Let $G$ be a graph database.

\paragraph{Any single path.}
Every single path in $G$ can be represented by a \pmr. Specifically, for a path
$\rho = v_0 e_1 v_1 e_2 v_2 \cdots e_n v_n$ in $G$, define
its \emph{canonical \pr} $R_\rho$ as $R_\rho = (N, E, \eta, \gamma, S, T)$ where
\begin{align*}
  N &= \{\bv_0,\ldots,\bv_n\} &
  \eta(\be_i) &= (\bv_{i-1},\bv_i) &\hspace{-1em}\text{for all } i \in [n]\\
  E &= \{\be_1,\ldots,\be_n\} &
  \gamma(\be_i) &= \be_i &\hspace{-1em}\text{for all } i \in [n] \\
  S &= \{\bv_0\},\ T=\{\bv_n\}&
  \gamma(\bv_j) &= v_j &\mathllap{\text{for all } j \in [0,n]}
\end{align*}
Then $\mpaths(R_\rho) = \spaths(R_\rho) = \{\rho\}$. Notice that we construct all
nodes $\bv_i$ and edges $\be_i$ in the \pr to be pairwise
distinct, while this is not necessarily the case for the nodes $v_i$ and edges
$e_i$ in $\rho$ (e.g., when $\rho$ has loops). This pairwise distinctness is necessary to ensure that $\mpaths(R_\rho)$ and $\spaths(R_\rho)$ are exactly the singleton $\{\rho\}$: had we simply taken $R_\rho$ to be the subgraph of $G$ induced by $\rho$ then, if $\rho$ contains loops, both $\mpaths(R_\rho)$ and $\spaths(R_\rho)$ would be infinite instead of the desired singleton.

\paragraph{Any finite multiset of paths.}
Let $M = \multileft \rho_1,\ldots,\rho_k \multiright$ be a finite multiset of paths in $G$.
For each path $\rho_i$, let $R_i$ be the canonical  \pr of $\rho_i$ and assume
w.l.o.g.\ that the sets of nodes and edges of these representations are pairwise
disjoint. Define the \emph{canonical \pr} $R_M$ of $M$ to be the disjoint union
$R_1 \sqcup \dots \sqcup R_n$ of the individual canonical \prs. Then $\mpaths(R_M) = M$. In the special case where every path in $M$ occurs only once, and $M$ is hence a set of
paths, then we also have that $\spaths(R_M) = M$.  

\begin{proposition}\label{prop:naive-path-representation}
  Let $M$ be a finite multiset of paths in $G$. Then there exists a path
  representation of $G$ that represents $M$.
\end{proposition}

The reader may wonder about which
infinite multisets of paths in $G$ can be represented by \pmrs. It turns out
that these are precisely the regular multisets, i.e., the multisets $M$ such
that $\set(M)$ is a regular language (i.e., there exists an NFA $A$ such that
$L(A) = \set(M)$) and $M(\rho)$ is the number of accepting runs of $A$
on $\rho$.

\subsection{Minimizing \pmrs}
\label{subsec:minim-complexity}

Reconsider the graph database of Figure~\ref{fig:2n-shortest-in-graph}, which
has $2^n$ paths from node $A$ to node $B$. Observe that, if we were to represent
this set of paths by means of the canonical \pr described in
Section~\ref{sec:basic-repr}, we end up with a representation that has $2^n$
elements (each containing $2n+1$ nodes) and is therefore of size
$\Omega(2^n)$. In Example~\ref{ex:succinct} we already noted, however, that
there exists an equivalent \pr that has only $3n+1$
nodes. The observatin that representations of the same path multiset may have
wildly varying sizes motivates a number of interesting computational tasks,
which we next phrase as decision problems.

\decisionproblem{\linewidth}{\pr Multiset Equivalence}
{Two
  \prs $R_1$ and $R_2$}{Is $\mpaths(R_1) = \mpaths(R_2)$?}

\decisionproblem{\linewidth}{\pr Multiset Minimization}{A \pmr $R$ and a number $k\in \N$}{Is there a \pr
  $R'$ with at most $k$ nodes and edges such that
  $\mpaths(R') = \mpaths(R)$?}

For each of the problems, we can also consider a ``Set'' version, in which we
ask the same questions but consider $\spaths$ instead of $\mpaths$ everywhere.

\begin{toappendix}
  To establish complexity bounds for PMR (multi)set Equivalence and Minimization, we observe that
  these problems are closely related to well known problems for NFAs. An NFA
  $A=(Q,\Sigma,\Delta,I,F)$ consists of a finite set of states $Q$, a finite
  alphabet $\Sigma$, a transition function $\Delta \colon Q \times \Sigma \to
  2^Q$, a set of initial states $I$ and a set of accepting states $F$.

  A \emph{path} in an NFA $A$ is a sequence
  \[
    \rho = q_0a_1q_1a_2q_2\dots a_nq_n
  \]
  with $n \geq 0$, $q_i \in Q$, $a_i \in \Sigma$, and $q_i \in
  \Delta(q_{i-1},a_1)$. We denote by $\paths(A)$ all paths in $A$ that start in
  a state $q_0 \in I$ and end in a state $q_n \in F$. For a given path $\rho$ we
  denote the corresponding word $a_1 \dots a_n$ by $\mathsf{Word}(\rho)$. We
  define the sets $\swords(A)$ and multisets $\mwords(A)$ of words accepted by
  an NFA as follows:
  \begin{align*}
    \swords(A) & := \{ a_1 \dots a_n \mid \exists \rho \in \paths(A).\; \mathsf{Word}(\rho) = a_1\dots a_n\}, \\
    \mwords(A) & := \multileft a_1 \dots a_n \mid \exists \rho \in \paths(A).\; \mathsf{Word}(\rho) = a_1\dots a_n \multiright . 
  \end{align*}

  We now establish the relatioship between \prs and NFAs. We interpret a \pr
  $R=(N,E,\eta,\gamma,S,T)$ over a database $G=(N_G,E_G,\allowbreak \eta_G,\allowbreak \lambda_G)$ as an
  NFA $A(R)=(N,E_G,\Delta,S,T)$, with \[\Delta \colon (v,e) \mapsto \{v' \mid
    \eta(e)= (\gamma(v),\gamma(v')) \} .\] That is the alphabet of the NFA is
  the set of edges of the database. The language (under set or multiset
  semantics) of $A(R)$ is the (multi)set of paths represented by $R$. We like to
  stress that in this relationship, we do not care about the nodes and labels of
  the database, as these are given implictly by the functions $\eta_g$ and
  $\lambda_G$. Also there is the purely formal difference that a string accepted
  by an NFA only consists of the labels and does not include information about
  the states.

  For the other direction, we can interpret an NFA $A=(Q,\Sigma,\Delta,I,F)$ as
  a \pr $R(A)=(Q,E,\eta,\gamma,I,F)$ with $E=\{(p,a,q) \mid q \in
  \Delta(p,a)\}$, $\eta \colon (p,a,q) \mapsto a$, and $\gamma \colon (p,a,q)
  \mapsto (p,q)$. This \pr is valid for any database that has $\Sigma$ as its
  set of edges.

  We note that the relationship is so close that we can minimize \prs by
  minimizing the corresponding NFA and interpreting the result as \pr again.
\end{toappendix}

Our first results are that, while set equivalence for \prs is \pspace-complete,
multiset equivalence is tractable. We obtain these results by connecting \pmr equivalence to equivalence problems for nondeterministic finite automata (NFAs). In particular,
we \inConfVersion{can} show that PMR set equivalence closely corresponds to language equivalence of
NFAs, which is known to be \pspace-complete in the worst
case~\cite{MeyerS-focs72} but for which methods exist that work well in
practice~\cite{MayrC-popl13}.  Moreover, we \inConfVersion{can} show that PMR multiset equivalence
corresponds to \emph{path equivalence} of NFAs, which asks if every word has an
equal number of accepting runs in both automata.  This is a non-trivial problem
that has been shown to be in polynomial time under some side-conditions that are
fulfilled by \prs~\cite{Tzeng-ipl96,Schutzenberger-iandc61b}.
We thus obtain:
\begin{toappendix}
  \begin{observation}\label{obs:GtoNFA}
    Let $R=(N,E,\eta,\gamma,S,T)$ be a path representation for a database
    $G=(N_G,E_G,\eta_G, \lambda_G)$. Then
    \begin{align*}
      \spaths(R)&=\{ v_0e_1 \dots e_nv_n \mid
      e_1\dots e_n \in \swords(A(R)) \text{ and}\\  & & \mathllap{(v_{i-1},v_i)=\eta(e_i) \text{ for } i
      \in \{1,\dots,n\} \},}\\
      \mpaths(R)&=\multileft v_0e_1 \dots e_nv_n \mid
      e_1\dots e_n \in \mwords(A(R)) \text{ and}\\  & & \mathllap{(v_{i-1},v_i)=\eta(e_i) \text{ for } i
      \in \{1,\dots,n\} \multiright .}
    \end{align*}
  \end{observation}
\end{toappendix}

\begin{theoremrep}\label{theo:prme}\label{theo:prse}
  \begin{enumerate}[(a)]
  \item \pr Multiset Equivalence is in \ptime.
  \item \pr Set Equivalence is \pspace-com\-plete.
  \end{enumerate}
\end{theoremrep}
\begin{proof}
  We first prove (a).
  Let $R_1$ and $R_2$ be two path representations over some database $G$. By
  Observation~\ref{obs:GtoNFA}, it holds that $\mpaths(R_1)=\mpaths(R_2)$, if and only if
  $\mwords(A(R_1)) = \mwords(A(R_2))$, i.e. if and only if $A(R_1)$ and $A(R_2)$
  are equivalent under multiset semantics. This is also called the path
  equivalence problem of NFAs, which is in \ptime if the automata do not have
  $\varepsilon$-transitions~\cite{Tzeng-ipl96,Schutzenberger-iandc61b}.

  We now show (b) and start with the upper bound. Let $R_1$ and $R_2$ be two path representations
  over some database $G$. By Observation~\ref{obs:GtoNFA}, it holds that
  $\spaths(R_1)=\spaths(R_2)$, if and only if $\swords(A(R_1)) =
  \swords(A(R_2))$, i.e. if and only if $A(R_1)$ and $A(R_2)$ are equivalent.
  This is the well known equivalence problem for NFAs, which is
  \pspace-complete~\cite{MeyerS-focs72}.

  For the lower bound, we reduce from NFA equivalence. Let $A_1, A_2$ be two
  NFAs over some alphabet $\Sigma$. By Observation~\ref{obs:GtoNFA}, $A_1$ and
  $A_2$ are equivalent if and only if $R(A_1)$ and $R(A_2)$ are set equivalent.
  As we do not care about database nodes in NFA equivalence, we can use the
  database $G=(\{v\}, \Sigma, \eta_G, \lambda_G)$ that has a single node $v$ and
  a self-loop edge for each label in $\Sigma$, i.e., $\eta_G \colon a
  \mapsto (v,v)$.
\end{proof}

We feel that these results are encouraging because (1) \pr
Multiset Equivalence is tractable and (2) while \pr Set
Equivalence is not worst-case tractable,
there are methods for dealing with the problem in practice~\cite{MayrC-popl13}. In
fact, we will show better complexity for \pr set equivalence for
practical settings involving regular path query evaluation
(Proposition~\ref{prop:dfa-equivalence}).

The close correspondence between \prs and NFAs is also useful
for establishing the complexities of \pr Multiset Minimization
and \pr Set Minimization. Our proofs show
that methods for minimizing NFAs can be directly used for these problems.
\begin{theoremrep}
  \begin{enumerate}[(a)]
  \item \pr Multiset Minimization is \np-complete.
  \item   \pr Set Minimization is \pspace-complete.
  \end{enumerate}
\end{theoremrep}
\begin{proof}
  We first show (a). We give an \np algorithm for Multiset Minimization. Guess a \pr of the desired
  size and then check whether it is equivalent with the given \pr. The check can
  be done in polynomial time by Theorem~\ref{theo:prme}.

  For NP-hardness we reduce from the minimization problem of unambigous NFAs,
  whcih is \np-complete~\cite{JiangR-siamcomp93}. For an unambgious NFA $A$ it holds
  by definition that $\mwords(A) = \swords(A)$. We can thus minimize $A$ by
  minimizing $R(A)$ under multiset semantics (which is the same as set semantics
  in this case) and then interpret the resulting \pr $R'$ as NFA $A(R')$.

  We now show (b).
  Membership in \pspace follows from the trivial algorithm that guesses a \pr of
  the desired size and checks equivalence. The check can be done in polynomial
  space by Theorem~\ref{theo:prse}.

  For \pspace-hardness we reduce from NFA minimization, which is
  PSPACE-complete~\cite{StockmeyerM-stoc73}. Let $A$ be an NFA. We can compute a
  minimal equivalent NFA $A'$ by minimizing $R(A)$ under set semantincs and then
  interpreting the resulting \pr $R'$ as NFA $A(R')$.
\end{proof}
In fact, the connection between \pr Set Minimization and NFA minimization is so
strong that methods such as bisimulation minimization \cite{PaigeT-siamcomp87}
or methods for B\"uchi automata minimization that carry over to the case of NFAs
\cite{MayrC-popl13}, which are commonly used in Software Verification, can be
directly applied to \pmrs. Notice that bisimulation minimization always runs in polynomial
time, but does not necessarily return the minimal result.

\begin{toappendix}
\forLater{For completeness, we might want to add that Gramlich and Schnitger~\cite{GramlichS-jcss07} showed that a given NFA with $n$ states cannot be efficiently minimized within factor $o(n)$ unless P $=$ PSPACE. }
\end{toappendix}

 \section{Generalized Regular Path Queries}
\label{sec:grpqs}

Regular path queries are a crucial feature that sets graph query languages apart
from relational query languages, since they allow us to easily ask queries about
arbitrarily long paths in graphs. Furthermore, they are central in Cypher
\cite{cypher}, SQL/PGQ, and GQL \cite{GQL-industry}. Although regular path
queries have been studied in research for decades (e.g.,
\cite{CruzMW-sigmod87,MendelzonW-sicomp95,CalvaneseGLV-pods99,Barcelo-pods13,MartensT-tods19}), their incarnation in Cypher, SQL/PGQ, and GQL is
different: they now have the capability of returning entire paths instead of
just their endpoints. In this section, we introduce \emph{generalized regular path queries} (\grpqs) to
formalize this important extra feature.

\smallskip
\noindent \textbf{Regular languages, expressions, and automata.}  We 
recap some basics on regular expressions and regular languages. A set of
words (each word using symbols from our fixed set of labels $\bL$) is also called a
\emph{language}. A \emph{regular expression} is an expression of the form 
\begin{align*}
  \expr &::=  \varepsilon \mid a \mid \expr_1 \expr_2 \mid \expr_1 + \expr_2 \mid \expr^* .
\end{align*}
Here, $\varepsilon$ denotes the empty
word and $a$ ranges over symbols in $\bL$. The \emph{language} $\lang(\expr)$
of expression $\expr$ is defined a
usual~\cite{HopcroftUIntroduction07}. A
language $L$ is \emph{regular} if there exists a regular expression $\expr$ such
that $L = \lang(\expr)$.\footnote{Notice that we have expressions
  for all regular languages, except the emtpy language, which is typically not used in the
  context of RPQs.} Regular languages can equivalently be represented by
finite state automata. We assume basic familiarity with deterministic (DFA) and
non-deterministic finite automata (NFAs)~\cite{HopcroftUIntroduction07}, and
omit their formal definition. We say that an NFA $A$ is \emph{unambiguous} (UFA
for short) if it has at most one accepting run for every word. Every DFA is
unambiguous, but the converse is not necessarily true.
In what follows we will range over regular expressions by the meta-variable $\expr$ and over UFAs by the meta-variable $\ufa$. We write $\lang(\expr)$ and $\lang(\ufa)$ to denote the language of $\expr$ and $\ufa$, respectively.

\smallskip
\noindent \textbf{Generalized regular path queries.}  While classical \rpqs are
syntactically defined to be simply a relational-calculus-like atom $(x,L,y)$ of
endpoint variables $(x,y)$ and regular language $L$, we find it convenient for
the development that will follow in Sections~\ref{sec:answering-grpqs}--\ref{sec:gcrpqs} to develop \grpqs as a small algebraic
query language. Specifically, our syntax for \grpqs completely ignores binding
endpoints to endpoint variables, as this feature is unimportant for the
immediate results that follow. We will re-introduce such variables when we
consider conjunctive generalized regular path queries in
Section~\ref{sec:gcrpqs}.

Formally, a \emph{Generalized Regular Path Query} (\grpq)
is an expression $\varphi$ of the form
\begin{align*}
  \varphi & ::= L \mid \sigma_{U,V}(\varphi) \mid m(\varphi)  \\
  m & ::= \shortestmode, \allowbreak \inFullVersion{\leximode,\;}
      \simplemode, \trailmode
\end{align*}
Here, $L$ is regular language (possibly specified by a regular expression or
UFA), $U$ and $V$ are either a finite set of  node identifiers or the infinite set of all node identifiers,\footnote{Our main use of
  $\sigma_{U,V}(\varphi)$ will be to restrict the endpoints of a result of a
  subquery $\varphi$ to sets of nodes $U$ and $V$ that we have already computed
  elsewhere in the query plan. From a systems perspective, it helps to think of
  $U$ and $V$ as pointers to sets (or unary predicates on nodes) rather than the
  sets themselves.} and $m$ is a \emph{selector mode}. We will refer to sets of node identifiers like $U$ and $V$ that are either finite or the set of all nodes as \emph{node predicates}.

Intuitively, $L$ selects all paths that match $L$, whereas $\sigma_{U,V}$
restricts results to those for which the source and target endpoints belong to
$U$ and $V$, respectively, and $m$ restricts results to those paths that are
shortest,
simple, or trail. Formally, a \grpq
$\varphi$, when evaluated on a graph database $G$, evaluates to a path multiset
$\varphi(G)$ over $G$, inductively defined as follows. Let $\paths(G)$ denote the
(possibly infinite) set of all paths of $G$.
\begin{align*}
  L(G) & = \bag{ \rho \in \paths(G) \mid \lambda_G(\rho) \in L }, \\
  \sigma_{U,V}(\varphi)(G) & = \bag{\rho \in \varphi(G) \mid \src(\rho) \in U, \tgt(\rho) \in V}, \\
  m(\varphi)(G) & = m(\varphi(G)).
\end{align*}
Let $M$ be any multiset of paths of $G$. In the last line, the semantics of
selector mode $m$ is defined by
\begin{align*}
  \shortestmode(M) &= \{\rho \in M \mid \rho \text{ is a shortest path}\}, \\
  \inFullVersion{\leximode(M)&= \{\rho \in M \mid \rho \text{ is a shortest path}\}, \\}
  \simplemode(M) &= \{\rho \in M \mid \rho \text{ is simple}\}, \text{ and} \\
  \trailmode(M) &= \{\rho \in M \mid \rho \text{ is a trail}\},
\end{align*}
where a path $\rho$ is a \emph{shortest path}, if there exists no shorter path
from $\src(\rho)$ to $\tgt(\rho)$ in $M$, \inFullVersion{it is a \emph{radix
    shortest path}, if it is a shortest path and there exists no smaller path in
radix order from $\src(\rho)$ to $\tgt(\rho)$ in $M$,} it is \emph{simple}, if each node
appears at most once in $\rho$, and it is a \emph{trail}, if each edge occurs at
most once in $\rho$. Notice that $\shortestmode(M)$ can contain paths
  of different length, since we only remove paths $\rho$ from $M$ for which
  there are shorter paths from $\src(\rho)$ to $\tgt(\rho)$.

The operations supported in \grpqs correspond to path evaluation modes in the
upcoming GQL standard~\cite{GQL-industry} and the ones studied in the research
literature. Specifically, the unrestricted version $L$ corresponds to regular
path queries~\cite{CruzMW-sigmod87}, and can return an infinite amount of paths,
such as in Example~\ref{ex:intro}. The $\trailmode$ mode is supported by Cypher
\cite{neo4-manual} and GQL \cite{GQL-industry}. The $\simplemode$ mode is similar,
but reverses the role of nodes and edges, and has been studied in the
literature \cite{MendelzonW-sicomp95,BaganBG-pods13,MartensT-tods19}. Finally, $\shortestmode$ is
supported by many existing systems~\cite{Stardog,neo4-manual,MillenniumDB}, and
the GQL standard~\cite{GQL-industry}. For a theoretical study of $\shortestmode$
see~\cite{shortest}.

\paragraph{Unions of \grpqs.} Note in particular that $\varphi(G)$, as defined
above, is actually a set of paths (no paths occur multiple times). This changes once we consider unions of \grpqs. A \emph{union of \grpqs} (\ugrpq for short) is an expression given by the syntax
\begin{align*}
  \psi & ::= \varphi \mid \psi \uplus \psi,
\end{align*}
where $\varphi$ ranges over \grpqs, and $\uplus$ denotes multiset union.
Formally, the semantics of a \grpq $\psi$ on a graph database $G$ is given by
$(\psi_1 \ \psi_2)(G) = \psi_1(G) \uplus \psi_2(G)$. The multiplicity of a
path in $(\psi_1 \uplus \psi_2)(G)$ is hence the sum of its multiplicity in
$\psi_1(G)$ plus its multiplicity in $\psi_2(G)$.

Whenever convenient, in what follows, we will apply the operators of \ugrpqs
directly on path multisets. For example, for a PM $M$ we write $\sigma_{U,V}(M)$
for $\bag{\rho \in M \mid \src(\rho) \in U, \tgt(\rho) \in V}$.

\smallskip
\noindent \textbf{Grouped output of \grpqs.} A (U)\grpq hence computes a path
(multi)\allowbreak set. Note that the elements of a PM are unsorted, so there
does not need to be any relationship between one path and the next. Sometimes,
however, it is desirable for efficiency reasons to \emph{group} the elements of
a PM, on their source node, target node, or both. This is the case, for
instance, when we wish to answer aggregate queries such as ``compute, for each
source node, the number of paths originating in that node'', or ``compute, for
each pair of endpoints $(u,v)$ the number of paths between them''. We next
formalize the notion of grouped path multisets.
\begin{definition}
  A \emph{(source/target/pairwise) grouped path multiset} (GPM) over a graph $G$ is a partition $H$ of a path
  multiset $M$ into maximal multisets, such that the following
  condition is satisfied for each multiset $M' \in H$:
  \begin{itemize}
  \item \emph{source grouped}: for all $\rho,\rho' \in M'$: $\src(\rho)=\src(\rho')$.
  \item \emph{target grouped}: for all $\rho,\rho' \in M'$:
    $\tgt(\rho)=\tgt(\rho')$. 
  \item \emph{pairwise grouped}: for all $\rho,\rho' \in M'$: $\src(\rho)=\src(\rho')$ and $\tgt(\rho)=\tgt(\rho')$.
  \end{itemize}
\end{definition}
Notice that, if $H$ is source grouped, it is a collection of multisets such that, for each $\rho_1 \in M_1 \in H$
and $\rho_2 \in M_2 \in H$ with $M_1 \neq M_2$, then $\src(\rho_1) \neq
\src(\rho_2)$. (The other cases are analogous.)

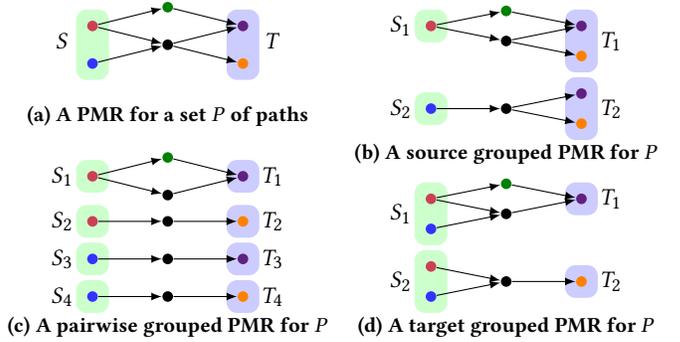
\begin{figure}[t]
  \begin{tikzpicture}
\node[circle,fill=BrickRed,inner sep=0pt,minimum size=4pt] (a) at (0,0) {};
    \node[circle,fill=green!50!black,inner sep=0pt,minimum size=4pt] (b) at (1,.25) {};
    \node[circle,fill=red!75!green!50!blue,inner sep=0pt,minimum size=4pt] (c) at (2,0) {};
    \node[circle,fill=blue!80,inner sep=0pt,minimum size=4pt] (d) at (0,-.5) {};
    \node[circle,fill=black,inner sep=0pt,minimum size=4pt] (e) at (1,-.25) {};
    \node[circle,fill=orange,inner sep=0pt,minimum size=4pt] (f) at (2,-.5) {};

    \node at (-.4,-.2) {$S$};
    \node at (2.4,-.2) {$T$};
    
		\path[-latex]
		(a) edge (b)
    (b) edge (c)
    (a) edge (e)
    (e) edge (c)
    (d) edge (e)
    (e) edge (f)
		;
    
    \begin{pgfonlayer}{background}
      \node[fill=green!20,inner
      sep=4pt,rectangle,rounded corners,fit=(a) (d)] {};
\node[fill=blue!20,inner
      sep=4pt,rectangle,rounded corners,fit=(c) (f)] {};
    \end{pgfonlayer}

\node at (1,-1.2) {\small \textbf{(a) A \pmr for a set $P$ of paths}};

\node (tr-anchor) at (4.5,0) {};
    \node[circle,fill=BrickRed,inner sep=0pt,minimum size=4pt] (tr-a1) at ($(tr-anchor)+(0,0)$) {};
    \node[circle,fill=green!50!black,inner sep=0pt,minimum size=4pt] (tr-b1) at ($(tr-anchor)+(1,.2)$) {};
    \node[circle,fill=red!75!green!50!blue,inner sep=0pt,minimum size=4pt] (tr-c1) at ($(tr-anchor)+(2,0)$) {};
    \node[circle,fill=black,inner sep=0pt,minimum size=4pt] (tr-e1) at ($(tr-anchor)+(1,-.2)$) {};
    \node[circle,fill=orange,inner sep=0pt,minimum size=4pt] (tr-f1) at ($(tr-anchor)+(2,-.4)$) {};

    \node[circle,fill=red!75!green!50!blue,inner sep=0pt,minimum size=4pt] (tr-c2) at ($(tr-anchor)+(2,-.9)$) {};
    \node[circle,fill=blue!80,inner sep=0pt,minimum size=4pt] (tr-d2) at ($(tr-anchor)+(0,-1.1)$) {};
    \node[circle,fill=black,inner sep=0pt,minimum size=4pt] (tr-e2) at ($(tr-anchor)+(1,-1.1)$) {};
    \node[circle,fill=orange,inner sep=0pt,minimum size=4pt] (tr-f2) at ($(tr-anchor)+(2,-1.3)$) {};

    \node at ($(tr-anchor)+(-.4,-.0)$) {$S_1$};
    \node at ($(tr-anchor)+(2.4,-.2)$) {$T_1$};
    \node at ($(tr-anchor)+(-.4,-1.1)$) {$S_2$};
    \node at ($(tr-anchor)+(2.4,-1.1)$) {$T_2$};

    \path[-latex]
    (tr-a1) edge (tr-b1)
    (tr-b1) edge (tr-c1)
    (tr-a1) edge (tr-e1)
    (tr-e1) edge (tr-c1)
    (tr-e1) edge (tr-f1)
    ;

    \path[-latex]
    (tr-e2) edge (tr-c2)
    (tr-d2) edge (tr-e2)
    (tr-e2) edge (tr-f2)
    ;

    \begin{pgfonlayer}{background}
      \node[fill=green!20,inner
      sep=4pt,rectangle,rounded corners,fit=(tr-a1)] {};
      \node[fill=green!20,inner
      sep=4pt,rectangle,rounded corners,fit=(tr-d2)] {};
      \node[fill=blue!20,inner
      sep=4pt,rectangle,rounded corners,fit=(tr-c1) (tr-f1)] {};
      \node[fill=blue!20,inner
      sep=4pt,rectangle,rounded corners,fit=(tr-c2) (tr-f2)] {};
    \end{pgfonlayer}
    
\node at ($(tr-anchor)+(1,-1.7)$) {\small \textbf{(b) A source grouped \pmr for $P$}};

\node (bl-anchor) at (0,-2) {};

    \node[circle,fill=BrickRed,inner sep=0pt,minimum size=4pt] (bl-a1) at ($(bl-anchor)+(0,0)$) {};
    \node[circle,fill=green!50!black,inner sep=0pt,minimum size=4pt] (bl-b1) at ($(bl-anchor)+(1,.25)$) {};
    \node[circle,fill=red!75!green!50!blue,inner sep=0pt,minimum size=4pt] (bl-c1) at ($(bl-anchor)+(2,0)$) {};
    \node[circle,fill=black,inner sep=0pt,minimum size=4pt] (bl-e1) at ($(bl-anchor)+(1,-.25)$) {};

    \node[circle,fill=blue!80,inner sep=0pt,minimum size=4pt] (bl-d2) at ($(bl-anchor)+(0,-1.6)$) {};
    \node[circle,fill=black,inner sep=0pt,minimum size=4pt] (bl-e2) at ($(bl-anchor)+(1,-1.6)$) {};
    \node[circle,fill=orange,inner sep=0pt,minimum size=4pt] (bl-f2) at ($(bl-anchor)+(2,-1.6)$) {};

    \node[circle,fill=BrickRed,inner sep=0pt,minimum size=4pt] (bl-a3) at ($(bl-anchor)+(0,-0.6)$) {};
    \node[circle,fill=black,inner sep=0pt,minimum size=4pt] (bl-e3) at ($(bl-anchor)+(1,-0.6)$) {};
    \node[circle,fill=orange,inner sep=0pt,minimum size=4pt] (bl-f3) at ($(bl-anchor)+(2,-0.6)$) {};

    \node[circle,fill=red!75!green!50!blue,inner sep=0pt,minimum size=4pt] (bl-c4) at ($(bl-anchor)+(2,-1.1)$) {};
    \node[circle,fill=blue!80,inner sep=0pt,minimum size=4pt] (bl-d4) at ($(bl-anchor)+(0,-1.1)$) {};
    \node[circle,fill=black,inner sep=0pt,minimum size=4pt] (bl-e4) at ($(bl-anchor)+(1,-1.1)$) {};

    \node at ($(bl-anchor)+(-.4,-.0)$) {$S_1$};
    \node at ($(bl-anchor)+(2.4,-.0)$) {$T_1$};
    \node at ($(bl-anchor)+(-.4,-.6)$) {$S_2$};
    \node at ($(bl-anchor)+(2.4,-.6)$) {$T_2$};
    \node at ($(bl-anchor)+(-.4,-1.1)$) {$S_3$};
    \node at ($(bl-anchor)+(2.4,-1.1)$) {$T_3$};
    \node at ($(bl-anchor)+(-.4,-1.6)$) {$S_4$};
    \node at ($(bl-anchor)+(2.4,-1.6)$) {$T_4$};

    \path[-latex]
		(bl-a1) edge (bl-b1)
    (bl-b1) edge (bl-c1)
    (bl-a1) edge (bl-e1)
    (bl-e1) edge (bl-c1)
		;

    \path[-latex]
		(bl-d2) edge (bl-e2)
    (bl-e2) edge (bl-f2)
		;

    \path[-latex]
		(bl-a3) edge (bl-e3)
    (bl-e3) edge (bl-f3)
		;

    \path[-latex]
		(bl-d4) edge (bl-e4)
    (bl-e4) edge (bl-c4)
		;

    \begin{pgfonlayer}{background}
      \node[fill=green!20,inner
      sep=4pt,rectangle,rounded corners,fit=(bl-a1)] {};
      \node[fill=green!20,inner
      sep=4pt,rectangle,rounded corners,fit=(bl-a3)] {};
      \node[fill=green!20,inner
      sep=4pt,rectangle,rounded corners,fit=(bl-d4)] {};
      \node[fill=green!20,inner
      sep=4pt,rectangle,rounded corners,fit=(bl-d2)] {};

\node[fill=blue!20,inner
      sep=4pt,rectangle,rounded corners,fit=(bl-c1)] {};
      \node[fill=blue!20,inner
      sep=4pt,rectangle,rounded corners,fit=(bl-c4)] {};
      \node[fill=blue!20,inner
      sep=4pt,rectangle,rounded corners,fit=(bl-f2)] {};
      \node[fill=blue!20,inner
      sep=4pt,rectangle,rounded corners,fit=(bl-f3)] {};

    \end{pgfonlayer}
    
\node at ($(bl-anchor)+(1,-2)$) {\small \textbf{(c) A pairwise grouped \pmr for $P$}};

\node (br-anchor) at (4.5,-2.3) {};
    \node[circle,fill=BrickRed,inner sep=0pt,minimum size=4pt] (br-a1) at ($(br-anchor)+(0,0)$) {};
    \node[circle,fill=green!50!black,inner sep=0pt,minimum size=4pt] (br-b1) at ($(br-anchor)+(1,.2)$) {};
    \node[circle,fill=red!75!green!50!blue,inner sep=0pt,minimum size=4pt] (br-c1) at ($(br-anchor)+(2,0)$) {};
    \node[circle,fill=blue!80,inner sep=0pt,minimum size=4pt] (br-d1) at ($(br-anchor)+(0,-.4)$) {};
    \node[circle,fill=black,inner sep=0pt,minimum size=4pt] (br-e1) at ($(br-anchor)+(1,-.2)$) {};

		\path[-latex]
		(br-a1) edge (br-b1)
    (br-b1) edge (br-c1)
    (br-a1) edge (br-e1)
    (br-e1) edge (br-c1)
    (br-d1) edge (br-e1)
;
    
    \node[circle,fill=BrickRed,inner sep=0pt,minimum size=4pt] (br-a2) at ($(br-anchor)+(0,-.9)$) {};
\node[circle,fill=blue!80,inner sep=0pt,minimum size=4pt] (br-d2) at ($(br-anchor)+(0,-1.3)$) {};
    \node[circle,fill=black,inner sep=0pt,minimum size=4pt] (br-e2) at ($(br-anchor)+(1,-1.1)$) {};
    \node[circle,fill=orange,inner sep=0pt,minimum size=4pt] (br-f2) at ($(br-anchor)+(2,-1.1)$) {};

    \node at ($(br-anchor)+(-.4,-.2)$) {$S_1$};
    \node at ($(br-anchor)+(2.4,-.0)$) {$T_1$};
    \node at ($(br-anchor)+(-.4,-1.1)$) {$S_2$};
    \node at ($(br-anchor)+(2.4,-1.1)$) {$T_2$};

		\path[-latex]
(br-a2) edge (br-e2)
(br-d2) edge (br-e2)
    (br-e2) edge (br-f2)
		;

    \begin{pgfonlayer}{background}
      \node[fill=green!20,inner
      sep=4pt,rectangle,rounded corners,fit=(br-a1) (br-d1)] {};
      \node[fill=green!20,inner
      sep=4pt,rectangle,rounded corners,fit=(br-a2) (br-d2)] {};

\node[fill=blue!20,inner
      sep=4pt,rectangle,rounded corners,fit=(br-c1)] {};
      \node[fill=blue!20,inner
      sep=4pt,rectangle,rounded corners,fit=(br-f2)] {};
    \end{pgfonlayer}

\node at ($(br-anchor)+(1,-1.7)$) {\small \textbf{(d) A target grouped \pmr for $P$}};

  \end{tikzpicture}
  \caption{Grouped \pmrs for the same set of paths\label{fig:grouped-reps}.}
\end{figure}

Let $M$ be a PM over a graph $G$. We define the following grouping
operators on $M$, which return a source grouped, target grouped, and pairwise
grouped GPM, respectively.
\begin{align*}
  \group{\src}(M) & = \{ \sigma_{\{\src(\rho)\},N_G}(M) \mid \rho \in M \},\\
    \group{\tgt}(M) & = \{ \sigma_{N_G,\{\tgt(\rho)\}}(M) \mid \rho \in M \},\\
    \group{\src,\tgt}(M) & = \{ \sigma_{\{\src(\rho)\},\{\tgt(\rho)\}}(M) \mid \rho \in M \}.
\end{align*}
We refer to Figure~\ref{fig:grouped-reps} for a visualization of the different groupings. (The
figure illustrates how we can use \pmrs for representing the different groups,
but may be helpful here nevertheless.)

We also introduce grouping at the query language level, and define a
\emph{grouped} \ugrpq to be an expression of the form $\group{S}(\psi)$ with
$\psi$ a \ugrpq and $S$ a non-empty subset of $\{\src, \tgt\}$. The semantics
of grouped \grpqs is the obvious one:
$\group{S}(\psi)(G) = \group{S}(\psi(G))$.

\smallskip
\noindent \textbf{Tabular output of (grouped) \ugrpqs.}  A (U)\grpq hence computes a path (multi)\allowbreak set, and a
grouped \ugrpq computes a grouped path multiset.

GQL, SQL/PGQ, and Cypher represent path multisets by means of a relational table
such as the one illustrated in Table~\ref{tab:tabularoutput-a}. To refer to this
representation, for a PM $M$, we write $\tab(M)$ for the table containing the
tuples $(\src(\rho), \tgt(\rho), \rho)$ for each $\rho \in M$. As such,
\[\tab(\psi(G)) = \bag{ (\src(\rho),\tgt(\rho), \rho) \mid \rho \in \psi(G) }\;.\]
We introduce a similar relational table representation on grouped PMs, and define
\begin{align*}
  \tab(\group{\src}(M)) & = \{ (\src(M'), M') \mid M' \in \group{\src}(M)) \} \;  \\
  \tab(\group{\tgt}(M)) & = \{ (\tgt(M'), M') \mid M' \in \group{\tgt}(M) \} \;  \\
  \tab(\group{\src,\tgt}(M)) & = \{ \left(\src(M'),\tgt(M'),M'\right) \mid M' \in \group{\src,\tgt}(M) \} \;  
\end{align*}
Here, we write $\src(M')$ (resp. $\tgt(M')$) for the unique source node (resp. target node) shared by all paths in $M'$.

It is important to stress  the difference between $\tab(\psi(G))$ and   $\tab(\group{\src,\tgt}(\psi(G)))$: the former has one tuple per path in $\psi(G)$, while the latter has one tuple per group in $\group{\src,\tgt}(\psi(G))$; the third component of that latter tuple is itself a path multiset.
To illustrate, Table~\ref{tab:tabularoutput-b} shows the pairwise-grouped tabular
representation for the \textsf{Transfer}-trails of
Figure~\ref{fig:propertygraph}, while Table~\ref{tab:tabularoutput-a} shows the ungrouped tabular representation.

\smallskip
\noindent \textbf{\pmr output for (grouped) \ugrpqs.}  Our interest in this paper
is in using \pmrs for succinctly representing the outputs of \ugrpqs. In this
respect, we say that a \pmr $R$ represents the output of \ugrpq $\psi$ on
graph $G$ if it represents $\psi(G)$.

Similarly to how \pmrs represent PMs, we introduce \emph{grouped \pmrs} to
represent grouped PMs. Concretely, a \emph{grouped \pmr} is a finite set
$S = \{R_1,\dots, R_k\}$ of \pmrs, such that $\mpaths(R_i)$ and $\mpaths(R_j)$
are disjoint, for every $i \not = j$. A grouped \pmr \emph{represents} a 
grouped PM $H$ if $H = \{\mpaths(R_1), \dots, \mpaths(R_k)\}$.

Figure~\ref{fig:grouped-reps} contains a \pr of five paths and
illustrates different groupings of the set of paths. We use six different colors to
show the six different nodes in $G$ under the image of $\gamma$.

 \section{Answering \ugrpqs}
\label{sec:answering-grpqs}

We now explore how to compute (grouped) PM representations for
answers of (grouped) \ugrpqs (in Section \ref{sec:rpq-evaluation}) and then discuss how to use PM representations to compute the tabular ouput of (grouped) \ugrpqs (Section~\ref{sec:computing-outputs-from-prs}), as well as graph projections (Section~\ref{sec:graphprojections-complexity}). We also explore related problems, such as counting the number of paths in a PMR, and drawing finite samples (Section~\ref{sec:computing-outputs-from-prs}).

\paragraph*{Model of computation.} To analyze the complexity of our algorithms,
we assume a RAM model of computation where the space used by node and edge ids,
as well as integers, the time of arithmetic operations on integers, and the
time of memory lookups are all $\bigo(1)$. We further assume that hash tables
have $\bigo(1)$ access and update times while requiring linear space. While it
is well-known that real hash table access is $\bigo(1)$ \emph{expected} time and
updates are $\bigo(1)$ \emph{amortized} time, complexity results that we establish
for this simpler model can be expected to translate to average (amortized)
complexity in real-life implementations~\cite{DBLP:books/mg/CormenLRS01}.

Throughout the rest of the paper, we assume an adjacency-list representation of
graph databases and \pmrs. As such, given a node $u$, it takes $\bigo(1)$ time
to retrieve the list of outgoing edges, while given an edge, it takes $\bigo(1)$
time to retrieve its endpoints. Retrieving the label of an edge is also
$\bigo(1)$, and the same holds for retrieving the value of the homomorphism
$\gamma$ for a node or edge in a \pmr.

\subsection{Computing Path Multiset Representations}
\label{sec:rpq-evaluation}\label{sec:computing-prs}

We first show how to compute a PMR for $\varphi(G)$ when $\varphi$ is a regular
language $L$. We will focus on the case where $L$ is given as an unambiguous
automaton $\ufa$. In practice, regular languages are always given as a regular
expression and, in theory, an exponential blow-up may occur when converting a
regular expression to a UFA. However, we inspected the regular expressions in
the query logs of \cite{BonifatiMT-www19,BonifatiMT-vldbj20}, with over
558 million SPARQL queries for Wikidata and DBpedia, containing 55 million RPQs, and
we noticed that for none of these expressions such a blow-up actually occurs:
the conversion is linear-time, even to a DFA. Our focus on UFAs is hence
reasonable. In what follows, if $\varphi$ is a \grpq, we write $\varphi = \ufa$
to indicate that $\varphi$ is of the form $L$, where the language $L$ is given by $\ufa$.

The fundamental notion that underlies our construction for representing
$\varphi(G)$ when $\varphi = \ufa$ is the
\emph{product} between a graph database and $\ufa$, which is defined as follows.
\begin{definition}[Graph product]\label{def:graph-product}
  Assume given an unabmiguous automaton $\ufa = (Q,\Sigma,\Delta,I,F)$, where
  $Q$ is the set of UFA states, $\Sigma \subseteq \bL$ is its set of used
  labels, $\Delta \subseteq Q \times \Sigma \times Q$ the set of
  transitions\footnote{Without loss of generality, we do not use
    $\varepsilon$-transitions.}, $I \subseteq Q$ is the set of initial states, and
  $F \subseteq Q$ the set of final states. Let $G = (N_G,E_G,\eta_G,\lambda_G)$ be a graph database. Then the
  \emph{product of $G$ and $\ufa$}, denoted as $G \times \ufa$, is the PM representation
  over $G$ defined as
\begin{itemize}
\item $N = N_G \times Q$
\item $E = \big\{\big(e,(q_1,a,q_2)\big) \in E_G \times \Delta \mid a=\lambda_G(e)\big\}$
\item $\eta((e,d)) = \big((v_1,q_1),(v_2,q_2) \big)$ such that
  \begin{itemize}
  \item $e$ is from $v_1$ to $v_2$ in $G$ and
  \item $d = (q_1,a,q_2)$, where $a = \lambda_G(e)$,
  \end{itemize}
\item $\gamma((v,q)) = v$,  $\gamma((e,d))= e$,
\item $S = N_G \times I$, and
\item $T =  N_G \times F$.
\end{itemize}
We will denote by $\trim(G \times \ufa)$ the subgraph of $G \times \ufa$ that is obtained by removing all nodes and
 edges that do not participate in a path from $S$ to $T$ in $G \times \ufa$. As
 such, $\trim(G \times \ufa)$ is a trim path multiset representation.
\end{definition}

Trimmed graph products provide a convenient way to obtain PM representations for
GRPQs of the form $\varphi = \ufa$. Indeed, we can show that $\trim(G \times \ufa)$ represents the set $\varphi(G) = \bag{ \rho \in \paths(G) \mid \lambda_G(\rho) \in \lang(\ufa)}$ of all $\ufa$-matched paths in $G$.

\begin{theoremrep}\label{theo:product-is-a-path-representation}
  Let $G$ be a graph database and let $\varphi = \ufa$ be a GRPQ. Then both
  $G \times \ufa$ and $\trim(G\times \ufa)$ are \pmrs of
  $\varphi(G)$, computable in linear time combined complexity
  $\bigo(\size{\varphi}\size{G})$.
\end{theoremrep}
\begin{proof}
  We first show correctness, by establishing that
  \begin{equation}
    \label{eq:1}
   \mpaths(G \times \ufa) = \varphi(G). 
  \end{equation}
  It then follows that also $\trim(G \times \ufa)$ represents $\varphi(G)$,
  because $\mpaths(\trim(G \times \ufa)) = \mpaths(G \times \ufa)$.

  To establish \eqref{eq:1}, we first observe that if $\rho$ and $\rho'$ are two
  paths from $S$ to $T$ in $G \times \ufa$ with $\gamma(\rho) = \gamma(\rho')$
  (i.e., $\rho$ and $\rho'$ encode the same path in $G$), then $\rho =
  \rho'$. In other words: the multiset
  $\mpaths(G \times \ufa) =\multileft \gamma(\rho) \mid \rho \text{ a path from
  } S \text{ to } T \text{ in } R \multiright$ is actually a set. This is
  because $\ufa$ is unambiguous. Indeed, assume for the purpose of obtaining a
  contradiction that $\gamma(\rho) = \gamma(\rho')$ but $\rho \not = \rho'$. Let
  \begin{align*}
    \rho & = (v_0,q_0) (e_1, d_1) (v_1, q_1) \dots (v_n, q_n) \\
    \rho' & = (v'_0,q'_0) (e'_1, d'_1) (v'_1, q'_1) \dots (v'_m, q'_m)
  \end{align*}
  where $v_0, \dots, v_n, v'_0, \dots, v'_m$ are nodes in $G$; $q_0,\dots, q_n, q'_0, \dots q'_m$ are states in $\ufa$ with $q_0 \in I$ and $q'_0 \in I$  initial states, $q_n \in F$ and $q'_n \in F$ final states, and $d_1, \dots, d_n, d'_1, \dots, d'_m$ transitions in $\ufa$. Because $\gamma(\rho) = \gamma(\rho')$ it follows that $m = n$, and $v_i = v'_i$ and $e_i = e'_i$ for $1 \leq i \leq m$. Because $\rho \not = \rho'$, however, the difference must be in the states or transitions. But then,
  $q_0, d_1, q_1, \dots, q_n $ and $q'_0, d'_1, q'_1, \dots, q'_n$ are two different accepting runs in \ufa of the word $\lambda_G(e_1) \dots \lambda_G(e_n)$, which contradicts our assumption that $\ufa$ is unambiguous.

  We remind the reader that the right-hand side  of \eqref{eq:1}, $\varphi(G)$, is also a set. This is because $\varphi(G) = \bag{ \rho \in \paths(G) \mid \lambda_G(\rho) \in \lang(\ufa)}$ and $\paths(G)$ itself is a set. 

  To establish \eqref{eq:1} let $\gamma(\rho)$ be an arbitrary element of $\mpaths(G \times \ufa)$, for some path $\rho =  (v_0,q_0) (e_1, d_1) (v_1, q_1) \dots (v_n, q_n)$ from $S$ to $T$ in $G \times \ufa$.  We show that $\gamma(\rho)$  is also in $\varphi(G)$. By definition of the product, we know that $\gamma(\rho) = v_0 e_1 v_1 \dots e_n v_n$ is a path in $G$. Moreover, again by definition, the sequence of edge labels $\lambda_G(e_1) \lambda_G(e_2) \dots \lambda_G(e_n)$ of the path $\gamma(\rho) = v_0 e_1 v_1 \dots e_n v_n$ is accepted by $\ufa$. Therefore, $\gamma(\rho)$ is a path in $G$ that matches $\lang(\ufa)$. Hence, $\gamma(\rho) \in \varphi(G)$.

  Conversely, let 
  $\rho'= v_0 e_1 v_1 \dots e_n v_n$ be a path in $\varphi(G)$.  That is, $\rho'$
  is in the right-hand side of \eqref{eq:1}. We show that also
  $\rho' \in \mpaths(G \times \ufa)$. By definition of the semantics of GRPQs,
  we know that $\lambda_G(\rho')$ is in $\lang(\ufa)$. There hence exists an accepting
  run $q_0 d_1 q_1 d_2 \dots d_n q_n$ of $\ufa$ on $\lambda_G(\rho')$. This
  implies that $\rho := (v_0,q_0) (e_1, d_1) (v_1, q_1) \dots (v_n, q_n)$ is a
  path from $S$ to $T$ in $G \times \ufa$ and hence that
  $\gamma' = \gamma(\rho) \in \mpaths(G \times \ufa)$.

We next discuss the complexity. Clearly, $G \times \ufa$ can be computed in time $\bigo(\size{q}\size{G})$. Computing $\trim(G \times \ufa)$ by  removing all nodes and edges that do not participate in a path from $S$ to $T$ in $G \times \ufa$ can then straightforwardly be done in $\bigo(\size{G \times \ufa}) = \bigo(\size{q}\size{G})$ by a variant of depth-first search (DFS), assuming an adjacency-list representation of $G \times \ufa$:
  \begin{itemize}
  \item Initially, mark all nodes and edges of $G \times \ufa$ as useless. If we
    assume that the ``mark'' of a node or edge is stored together with the node
    or edge, this is just a scan of $G \times \ufa$ and hence takes time
    $\bigo(\size{G \times \ufa}) = \bigo(\size{q}\size{G})$.
  \item Perform DFS on $G \times \ufa$, starting from the nodes in $S$.  DFS takes time $\bigo(\size{G \times \ufa}) = \bigo(\size{q}\size{G})$.
  \item During the DFS, keep track of the path $\rho$  traversed from the source node in $S$ to the current node. 
    Whenever the DFS reaches a node in $T$, mark all nodes in $\rho$ as useful. Because in a DFS we never traverse a node twice, $\rho$ can be of size at most the number of nodes in $G \times \ufa$, i.e., of size $\bigo(\size{q}\size{G})$, and marking them hence takes time $\bigo(\size{q}\size{G})$.
  \item During the DFS, whenever we try to traverse an edge from the current node $n$ that leads to an already visited node $v$ that is useful, mark all nodes in the current path $\rho$ leading up to $n$ as useful.
  \item When the DFS is done, iterate over all edges of $G \times \ufa$ and mark an edge useful if it connects two useful nodes.  Again, this is $\bigo(\size{q}\size{G})$.
  \item Remove all useless nodes and edges. This entails another iteration over the nodes (and their adjacency lists) and edges of $G \times \ufa$, and can hence be done in $\bigo(\size{q}\size{G})$
  \end{itemize}
  The result is a trimmed representation.
\end{proof}

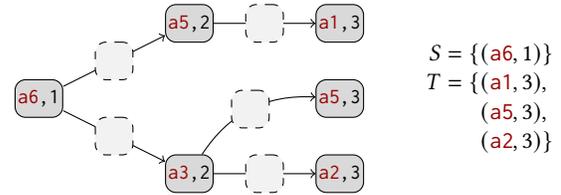
\begin{figure}
  \begin{tikzpicture}[initial text={}]
    \tikzstyle{every state}=[draw,rectangle,rounded corners,fill=white!85!black,minimum size=5mm, text=black, font=\ttfamily, inner sep=1pt]
    \tikzstyle{every node}=[font=\ttfamily]

\node[state] (a61) at (0,1) {\small \nodeid{a6},1};
    \node[state] (a32) at (2,0) {\small \nodeid{a3},2};
    \node[state] (a23) at (4,0) {\small \nodeid{a2},3};
    \node[state] (a52) at (2,2) {\small \nodeid{a5},2};
    \node[state] (a13) at (4,2) {\small \nodeid{a1},3};
    \node[state] (a53) at (4,1) {\small \nodeid{a5},3};
    
\path[->]
    (a61) edge node[state,dashed,fill=white!95!black] {} (a32)
    (a61) edge node[state,dashed,fill=white!95!black] {} (a52)
    (a52) edge node[state,dashed,fill=white!95!black] {} (a13)
    (a32) edge[bend left] node[state,dashed,fill=white!95!black] {} (a53)
    (a32) edge node[state,dashed,fill=white!95!black] {} (a23)
    ;

    \node at (6,1)
    {$\begin{array}{r@{\hspace{0mm}}l}
        S = \{ & (\nodeid{a6},1)\}\\
        T = \{ & (\nodeid{a1},3),\\
               & (\nodeid{a5},3),\\
               & (\nodeid{a2},3)\}
      \end{array}$};
  \end{tikzpicture}
  \caption{Illustration of the product construction..\label{fig:singleproduct}}  
\end{figure}

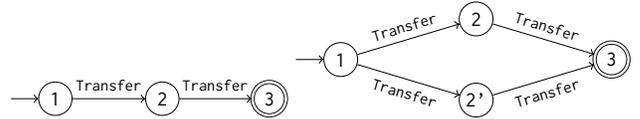
\begin{figure}[t] 
  \begin{minipage}[b]{.48\linewidth}
    \resizebox{.9\linewidth}{!}{
      \hspace{-4mm}
    \begin{tikzpicture}[initial text={}]
      \tikzstyle{every state}=[draw,minimum size=5mm, text=black, font=\ttfamily, inner sep=0pt]
      \tikzstyle{every node}=[font=\ttfamily]
      \def\x{1.6}
      \def\y{.6}
      
\node[state,initial] (1) at (\x*0,\y*0) {1};	
      \node[state] (2) at (\x*1,\y*0) {2};
      \node[state,accepting] (4) at (\x*2,\y*0) {3};
      
\path[->]
      (1) edge node [above,sloped] {\footnotesize Transfer} (2)
      (2) edge node [above,sloped] {\footnotesize Transfer} (4)
      ;
    \end{tikzpicture}}
\end{minipage}
  \begin{minipage}[b]{.48\linewidth}
    \resizebox{1\linewidth}{!}{
      \hspace{-8mm}
      \begin{tikzpicture}[initial text={}]
        \tikzstyle{every state}=[draw,minimum size=5mm, text=black, font=\ttfamily, inner sep=0pt]
        \tikzstyle{every node}=[font=\ttfamily]
        \def\x{2}
        \def\y{.6}
	
\node[state,initial] (1) at (\x*0,\y*0) {1};	
        \node[state] (2) at (\x*1,\y*1) {2};
        \node[state] (3) at (\x*1,\y*-1) {2'};
        \node[state,accepting] (4) at (\x*2,\y*0) {3};
      
\path[->]
        (1) edge node [above,sloped] {\footnotesize Transfer} (2)
        (2) edge node [above,sloped] {\footnotesize Transfer} (4)
        (1) edge node [below,sloped] {\footnotesize Transfer} (3)
        (3) edge node [below,sloped] {\footnotesize Transfer} (4)
        ;
      \end{tikzpicture}}
\end{minipage}
  \caption{Two automata for the language \textsf{Transfer}$\cdot$\textsf{Transfer}.}
	\label{fig:transfer-nfa}
      \end{figure}

We illustrate by means of the following example that the unambiguous property of
$\ufa$ in Theorem~\ref{theo:product-is-a-path-representation} is important for
the correctness of the construction. Specifically, it is needed to ensure
correct multiplicities of paths.  
\begin{example}\label{ex:canonicalPR}
  Consider the regular expressions $\expr_1 = \textsf{Transfer} \cdot
  \textsf{Transfer}$ and $\expr_2 = (\textsf{Transfer} \cdot \textsf{Transfer})
  + (\textsf{Transfer} \cdot \textsf{Transfer})$. Notice that $L(\expr_1) =
  L(\expr_2)$ and that $\expr_2$ is written in a ``non-optimal'' way.
  Nondeterministic automata that correspond to $\expr_1$ and $\expr_2$ are
  depicted in Figure~\ref{fig:transfer-nfa}: the left one is unambiguous (even
  deterministic) while the right one is not.
Figure~\ref{fig:singleproduct} illustrates a part of $\trim(G \times \ufa)$
  where $G$ is the graph from Figure~\ref{fig:propertygraph} and $\ufa$ is the
  left automaton from Figure~\ref{fig:transfer-nfa}, namely the part that is
  reachable from the node $(\nodeid{a6},1)$. The resulting \pmr represents three
  paths of length two in $G$, which means that three paths that match $\expr_1$
  start from $\nodeid{a6}$ in Figure~\ref{fig:propertygraph}. Notice that, if we
  would apply the same construction using the right NFA of
  Figure~\ref{fig:transfer-nfa}, the result would have two additional nodes
  $(\nodeid{a5},2')$ and $(\nodeid{a3},2')$, leading to 6 paths in $G$ (two
  copies of each path represented in Figure~\ref{fig:singleproduct}), which is incorrect.

 Obviously, both constructions are correct if multiplicities are not important (i.e., we are interested in the $\spaths$ semantics), but only the construction using the UFA has the correct
 multiplicities. \qed
\end{example}

\smallskip \noindent \textbf{Selection.}
We now consider \grpqs that involve the selection
operator $\sigma_{U,V}$.  Concretely, for a GRPQ $\varphi = \sigma_{U,V}(\ufa)$
and graph database $G$ we can obtain a PMR of $\varphi(G)$ by constructing
$G \times \ufa$, but trimming differently. In general, we observe that a more general way of trimming allows us to express $\sigma_{U,V}$ on arbitrary PMRs.
Concretely, let $R = (N,E,\eta,\gamma,S_R,T_R)$ be
an arbitrary PMR over a graph $G$ and let $U,V$ be node predicates. Denote by $\trim(R, U, V)$ the subgraph of $R$ that is obtained by
removing all nodes and edges in $R$ that do not participate in a path from
$S' := \{ u \in S_R \mid \gamma(u) \in U\}$ to
$T' := \{u \in T_R \mid \gamma(u) \in V\}$. \inConfVersion{We can show that}\inFullVersion{We show in the Appendix that}  $\trim(R, U, V)$ represents $\sigma_{U,V}(\mpaths(R))$, leading to the following theorem. For a node predicate $U$, let $\size{U}$ denote the cardinality of $U$ if $U$ is finite, and let it be $1$ if $U$ is the infinite set of all node identifiers\footnote{If $U$ is the set of all node identifiers, it corresponds to the predicate 'True', which is concisely represented in constant space, hence we set $\size{U} = 1$ in this case.}.

\begin{theoremrep}\label{theo:singleproduct}
  Let $R$ be a PMR on graph $G$. Let $U$ and $V$ be node predicates. Then $\trim(R, U, V)$ is a PMR of $\sigma_{U,V}(\mpaths(R))$, computable in linear time $\bigo(\size{R} + \size{U} + \size{V})$.
\end{theoremrep}
\begin{proof}
  By definition of $\trim(R, U, V)$ it follows that
  \begin{align*}
& \mpaths(\trim(R, U, V)))\\
    & = \multileft \gamma(\rho) \mid \rho \text{ path from } S_R \text{ to } T_R \text{ in } R,  \gamma(\src(\rho)) \in U, \gamma(\tgt(\rho)) \in V\multiright \\
& = \multileft \rho' \in \mpaths(R) \mid \src(\rho') \in U, \tgt(\rho') \in V  \multiright \\
& = \sigma_{U,V}(\mpaths(R))
  \end{align*}

  This establishes correctness. For the complexity, we note that trimming can be done in $\bigo(\size{R})$ time using a depth-first search similar to the proof of Theorem~\ref{theo:product-is-a-path-representation}. The only difference here is that instead of starting the DFS from nodes in $S$ and terminating in nodes in $T$, we first compute
  \begin{align*}
    S' &= \{ u \in S_R \mid \gamma(u) \in U \}, \\
    T' &= \{ u \in T_R \mid \gamma(u) \in V \}, 
  \end{align*}
  and trim using these sets.

Since $S_R$ and $T_R$ are subsets of the nodes in $R$, we can compute $S'$ and $T'$ by iterating over $S_R$ (resp. $T_R$), and for each such node $u$, check whether $\gamma(u) \in U$ (resp. $\gamma(u) \in V$) holds. If $U$ (or $V$) is the set of all nodes, this can be done in constant time $\bigo(1) = \bigo(\size{U})$. If $U$ (or $V$) is a finite set, this can be also be done in constant time, provided that before computing $S'$ (resp. $T$') we preprocess $U$  (resp. $V$) by building a hash table for it. In the RAM model, building such a hash table takes $\bigo(\size{U})$ time, after which lookups take $\bigo(1)$ time.

  The DFS from and to these sets are then as in Theorem~\ref{theo:product-is-a-path-representation}. As such, the whole trimming can be done in the claimed complexity.
\end{proof}

Consequently, we can evaluate \grpqs such as $\varphi = \sigma_{U,V}(\ufa)$ simply by computing $\trim(G \times \ufa, U, V)$ which, by Theorems~\ref{theo:product-is-a-path-representation} and \ref{theo:singleproduct}, can be done in linear time combined complexity $\bigo(\size{\varphi}\size{G} + \size{U} + \size{V})$.

\smallskip \noindent \textbf{Grouping.}
Since source grouped, target grouped, and pairwise grouped representations
of a PM $M$ can always be obtained by repeatedly computing
$\sigma_{U,V}(M)$ for different sets $U$ and $V$, we obtain the following
corollary from Theorem~\ref{theo:singleproduct}.

\begin{corollaryrep}
  \label{cor:grouped-all}
  Let $M$ be a PM on a graph $G$, represented by trim PMR $R$. Assume that $X$ is the
  set of all source nodes in $M$, i.e., $X = \{ \src(\rho) \mid \rho \in
  M\}$. Let $Y$ be the set of all target nodes in $M$, and let $XY$ be the set
  of all $(\src,\tgt)$ pairs of paths in $M$. We can then compute
  \begin{enumerate}
  \item a grouped representation of $\group{\src,\tgt}(M)$ in time $\bigo(\size{XY}\size{R})$;
  \item a grouped representation of $\group{\src}(M)$ in time 
$\bigo(\size{X}\size{R})$;
  \item a grouped representation of $\group{\tgt}(M)$ in time 
$\bigo(\size{Y}\size{R})$.
  \end{enumerate}
\end{corollaryrep}
\begin{proof}
To keep the notation consistent, we recall that $G = (N_G,E_G,\eta_G,\lambda_G)$, and that $R = (N_R,E_R,\eta_R,\gamma_R,S_R,T_R)$. Next we prove each item of the Corollary.

  (1) Recall a that grouped representation of the grouped PM
  $\group{\src,\tgt}(M)$ is a set $H=\{R_1,\dots, R_k\}$ of PMRs such that
  \begin{align*}
    \{ \mpaths(R_1), \dots, \mpaths(R_k) \} & =  \group{\src,\tgt}(M) \\
    & = \{ \sigma_{\{\src(\rho)\},\{\tgt(\rho)\}}(M) \mid \rho \in M \} \\ 
    & = \{ \sigma_{\{x\},\{y\}}(M) \mid (x,y) \in XY \}. 
  \end{align*}

We compute this pairwise grouped representation as follows. 
We first compute the set $XY$ from $R$ as follows:
\begin{itemize}
\item Initialize $XY$ to the empty set.
\item Compute the set $X$ in time $\bigo(\size{R})$, using the technique
  outlined in the correctness of point (2).
  \item For each node $x \in X$:
    \begin{itemize}
    \item compute the set of nodes
      $S_x = \{u \in S_R \mid \gamma_R(u) = x\}$. This can clearly be
      done in $\bigo(\size{S_R}) = \bigo(\size{R})$ time by one iteration over $S_R$.  (Here, we assume that from a node $u$ of $R$, we can compute $\gamma_R(u)$ in constant time, which is the case if $\gamma_R(u)$ is stored together with $u$.)
      \item Calculate, using DFS starting from $S_x$,  the set of all nodes $v \in T_R$ that are reachable
        from nodes in $S_x$. For each such $v$, add 
        $(\gamma_R(u), \gamma_R(v))$ to $XY$. DFS is in time linear in $R$.
    \end{itemize}
  \end{itemize}
  The total time spent computing $XY$ is hence
  $\bigo(\size{X}\size{R})$.
  
  This correctly computes $XY$ because $R$ represents $M$ and hence,
  \begin{align*}
    XY &= \{ (\src(\rho), \tgt(\rho)) \mid \rho \in M \} \\
       & = \{ (\src(\rho), \tgt(\rho)) \mid \rho \in \spaths(R) \} \\
       & = \{ (\gamma_R(u), \gamma_R(v)) \mid u \in S_R, v \in T_R, v \text{ reachable from } u \text{ in } R \}\\
       & = \{ (\gamma_R(u), \gamma_R(v)) \mid x \in X, u \in S_x, v \in T_R,\\ & \phantom{= \{ (\gamma_R(u), \gamma_R(v)) \mid\quad} v \text{ reachable from } u \text{ in } R \}.    
  \end{align*}
  Having computed $XY$, we next compute, for each $(x,y) \in XY$, a PMR for
  $\sigma_{\{x\},\{y\}}(M)$. Denote this representation by $R_{x,y}$. By
  Theorem~\ref{theo:singleproduct} each such representation can be computed in
  time $\bigo(\size{R} + \size{\{x\}} + \size{\{y\}}) = \bigo(\size{R})$. The
  output is now the set of PRMs $\{ R_{x,y} \mid (x,y) \in XY\}$. The total time
  spent is $\bigo(\size{X} \size{R} + \size{XY}\size{R})$. Because $\size{X} \leq \size{XY}$, this is $\bigo(\size{XY}\size{R})$, as desired.

\medskip

(2) Recall a that grouped
  representation of the grouped PM $\group{\src}(M)$ is a set
  $H=\{R_1,\dots, R_k\}$ of PMRs such that
  \begin{align*}
    \{ \mpaths(R_1), \dots, \mpaths(R_k) \} & =  \group{\src}(M) \\
    & = \{ \sigma_{\{\src(\rho)\}, N_G}(M) \mid \rho \in M \} \\ 
    & = \{ \sigma_{\{x\},N_G}(M) \mid x \in X \}. 
  \end{align*}

We compute this grouped representation as follows. We first compute $X$ from $R$ in time  $\bigo(\size{R})$ as follows:
\begin{itemize}
\item We represent $X$ by a hash table, which is initially empty.
\item Iterate over the set  $S_R$
  of $R$'s source nodes and add node $\gamma_R(u)$ to $X$, for each $u \in S_R$. By checking that $\gamma_R(u)$ is not already in $X$ prior to adding (which can be done in $\bigo(1)$ time in the RAM model) we ensure that no node in $X$ is represented twice. 

  (We assume here that calculating $\gamma_R(u)$ given node $u$ is $\bigo(1)$,
  which is the case if each node in $R$ carries its $\gamma_R$-mapped source
  node.)
\end{itemize}

This correctly computes $X$ because $R$ represents $M$ and hence,
  \begin{align*}
    X &= \{ (\src(\rho) \mid \rho \in M \} \\
       & = \{ (\src(\rho) \mid \rho \in \spaths(R) \} \\
       & = \{ \gamma_R(u) \mid u \in S_R \}.
  \end{align*}
Here, the last equality follows from the fact that $R$ is trim.

Having computed $X$, we next compute, for each $x \in X$, a PMR for $\sigma_{\{x\}, N_G}(M)$. Denote this representation by
$R_{x}$. By Theorem~\ref{theo:singleproduct} each such representation can be
computed in time $\bigo(\size{R})$ from $R$. The output is now the set of PMRs
$H = \{ R_x \mid x \in X\}$. The total time spent is
$\bigo(\size{R} + \size{X}\size{R}) =
\bigo(\size{X}\size{R})$, as desired. 
  
\newcommand{\rev}[1]{\ensuremath{{#1}^{-1}}}

\medskip
(3) To obtain our result, we define a \emph{reversal} operation on automata, path multisets, and path representations, as follows.

Let $\rev{\ufa}$ be the automaton obtained from $\ufa$ by simultaneously: (1) reversing the direction of all transitions of $\ufa$ (i.e., a transition $(q_1, a, q_2)$ becomes $(q_2, a, q_1)$), (2)  making the initial states final and (3) making the final states initial. Then  $\rev{\ufa}$ recognizes the reverse language of $\ufa$, i.e., $\lang(\rev{\ufa}) = \{ \rev{w} \mid w \in \lang(\ufa)\}$ where $\rev{w}$ denotes the string reversal of $w$. Let $\rev{G}$ be the graph database obtained from $G$ by reversing all edge directions in $G$ (i.e., if $\eta_G(e) = (n_1, n_2)$ then this becomes $(n_2, n_1)$). Let $\rev{M}$ be the path multiset on $\rev{G}$ obtained from $M$ by reversing each path $\rho = v_0e_1v_1e_2v_2 \cdots e_nv_n$ in $M$, i.e., replacing it by $\rev{\rho} = v_ne_nv_{n-1} \cdots e_1v_0$. Finally, let $\rev{R}$ be the PMR obtained from $R$ where all edge directions are reversed, source nodes become target nodes, and target nodes become source nodes.

It is not difficult to see that  computing a target grouped representation for $\group{\tgt}(M)$ is the same as computing a source  grouped representation $H$ of  $\group{\src}(\rev{M})$, and then reversing again each PMR of $H$. Because reversal of PMR $R$ (and the elements of $H$) is a linear time operations, the claimed complexity then follows immediately from point (2) proved above.
\end{proof}
We stress that the complexities given by Corollary~\ref{cor:grouped-all} are
attractive and, in a sense, optimal. Indeed, consider, for example, source
grouping $\group{\src}(M)$. There are $\size{X}$ groups in the resulting grouped
PM, and we hence need to represent every group by a PMR in a corresponding
grouped PMR. Corollary \ref{cor:grouped-all} tells us that a representation for
each such group can be obtained in linear time in the size of the original
representation $R$ of $M$. Since, without special preprocessing, we cannot even
read $R$ in less time, the resulting complexity is optimal.

\forLater{For pairwise grouped it might be possible to even show optimality conditioned on the matrix multiplication conjecture here. There is a link with the ICDT paper on the product construction of 2 years ago? Not sure if it is worth developing if we lack the space.}

\paragraph*{Shortest paths.}
We next turn our
attention to evaluating \grpqs that involve selector modes \inConfVersion{$m \in \{\shortestmode,
\allowbreak \simplemode,
\trailmode\}$.}\inFullVersion{$m \in \{\shortestmode, \allowbreak \leximode,
  \allowbreak \simplemode,
  \trailmode\}$.}
The next theorem shows that it is possible to apply
\inConfVersion{$\shortestmode$ as operation}\inFullVersion{$\shortestmode$ and $\leximode$ as operations}
on \prs with favorable
complexities. We will see later that $m = \simplemode$ and $m = \trailmode$
are more complex.

\begin{theoremrep}\label{thm:shortest-lexi-on-prs}
  Let $M$ be a PM over graph database $G$ and let $R$ be a trim \pmr
  representing $M$.  Let $k = \min(\size{S},\size{T})$ with $S$ and $T$
  the sets of source and target nodes of $R$, respectively. From $R$ we can
  compute \inConfVersion{a trim \pmr for $\shortestmode(M)$ in time
    $\bigo(k\size{R})$.}\inFullVersion{\begin{enumerate}
  \item a trim \pmr for $\shortestmode(M)$ in time $\bigo(k\size{R})$ and
  \item a trim \pmr for $\leximode(M)$ in time $\bigo(k\size{R}^2)$.
  \end{enumerate}}
\end{theoremrep}
\begin{proofsketch}
Recall that given a graph $G$ and a node $n$, using the standard breadth-first search (BFS) algorithm~\cite{DBLP:books/mg/CormenLRS01}, we can compute all the nodes reachable from $n$, as well as the distance of the shortest path to each node. By storing the predecessor used to reach the processed node, we can also reconstruct one shortest path witnessing this connection. It is not difficult to see that this algorithm can be extended, without modifying the complexity, by recording \emph{all} the predecessors of the processed node on any shortest path reaching this node~\cite{shortest}. This, in turn, allows to reconstruct all the shortest paths to any node reachable from $n$.

If we now assume that $S=\{n\}$ contains a single source node, we can run this modified BFS on the \pmr $R$ starting from $n$. This will contain enough information to allow further trimming $R$ in order to remove paths (under $\gamma$) which are not shortest in $G$, thus obtaining a \pmr for $\shortestmode(M)$. The complexity of this is the same as the complexity of BFS; namely, $\bigo(R)$. When $S$ contains multiple nodes, we simply repeat this process for each starting node, and take the disjoint union of the obtained \pmrs. Finally, by observing that the algorithm can be run in reverse, we can either use $S$, or $T$, to perform the construction, resulting in the desired complexity.
\end{proofsketch}
\begin{proof}
  \newcommand{\pred}{\ensuremath{\textit{pred}}}
  \newcommand{\preds}{\ensuremath{\textit{preds}}}
  Let $R=(N,E,\eta,\gamma,S,T)$. \inFullVersion{Let us begin with $\shortestmode(M)$.}
  We first illustrate how to compute
  a \pr for $\shortestmode(M)$ when $R$ contains only a single source node $n$
  (i.e., $S = \{n\}$).  It is well-known and well-documented (e.g.,
  \cite[section 22.2]{DBLP:books/mg/CormenLRS01}) that using standard
  breadth-first search (BFS) on $R$ starting from node $n$ one can compute for
  each node $m \in R$, the distance $\delta(n,m)$ from $n$ to $m$. Here the
  distance is defined as the number of edges traversed in the shortest path from
  $n$ to $m$ in $R$, where we take $\delta(n,m) = +\infty$ if there is no such
  path. Moreover, during this BFS one can record, for each node $m$, a
  \emph{predecessor node} $\pred(m)$. This predecessor node (when it is defined)
  has the property that (1) there is an edge from $\pred(m)$ to $m$ and (2) a
  shortest path from $n$ to $m$ is formed by a shortest path from $n$ to
  $\pred(m)$, followed by the edge from $\pred(m)$ to $m$. It is not difficult
  to see that, without modifying the complexity, we may instead record, during
  the BFS, the set $\preds(m)$ of all predecessor nodes of $m$ that appear in
  \emph{any} shortest path from $n$ to $m$: all nodes in $\preds(m)$ have an
  edge to $m$ and every shortest path (there can be multiple) from $n$ to $m$
  can be formed by first traversing a shortest path from $n$ to a node in
  $\preds(m)$ and then following the edge from that node to $m$. Moreover, every
  node in $\preds(m)$ appears in such a shortest path. For a detailed description of this version of BFS see e.g.~\cite{shortest}.

  It follows that this modified BFS can be used to compute a \pr for
  $\shortestmode(M)$ when $R$ has only one initial node $n$ (i.e. $S=\{n\}$):
  \begin{enumerate}
  \item Perform the modified BFS starting from $n$. This takes time $\bigo(\size{R})$.
  \item Compute the set $D$ of pairs
    $D = \{ (\gamma(m), \delta(n,m)) \mid m \in T\}$. This hence contains all
    nodes $\gamma(m)$ in the source graph $G$ that serve as an endpoint in a
    path from $n$ to $T$ in $R$. Computing $D$ takes time $\bigo(\size{T}) = \bigo(\size{R})$.
  \item Remove from $D$ all pairs $(\gamma(m), \delta(n,m))$ for which there
    exists node $m' \in T$ such that  $\gamma(m) = \gamma(m')$ and $\delta(n,m)
    > \delta(n,m')$. This can be done by iterating over $T$ and constructing a
    hash  table that groups on $\gamma(m)$ and keeps the shortest distance seen
    so far. In the RAM model, where hashing is constant-time, this hence takes $\bigo(\size{T})) = \bigo(\size{R})$. 

  \item Using  $D$ (now represented as a hash table that allow lookups on $\gamma(m)$ to find the minimal length $\delta(n,m)$ of any path in $R$ that starts in $\gamma(n)$ and ends in $\gamma(m)$), compute the subset $T'$ of $T$ defined by $T' = \{ m \in T \mid (\gamma(m), \delta(n,m)) \in D\}$. Now $T'$ contains, for each endpoint $\gamma(m')$ in the source graph $G$ that is reachable from $\gamma(n)$ by a path in $R$, only those encoding nodes $m$ in $T$ that participate in a shortest path from $\gamma(n)$ to $\gamma(m')$ in $R$. Again this step takes time $\bigo(\size{T}) = \bigo(\size{R})$.
  \item Use $\preds(m)$ to search backward, starting from $T'$ to $n$. Again this takes time $\bigo(\size{R})$.
  \item Remove from $R$ all nodes and edges that did not participate in this backward search. This can again be done in time $\bigo(\size{R})$ by marking all nodes and edges as useless before the backward search, marking nodes and edges traversed during the backward search as useful, and finally removing all useless nodes and edges. 
  \end{enumerate}
  The result is a trim path representation $R'$ on $G$, computed in time
  $\bigo(\size{R})$. It is clear that, by construction,
  \[ \mpaths(R') = \shortestmode(\mpaths(R)) = \shortestmode(M),\] as desired.

  To compute $\shortestmode(R)$ when $R$ has multiple source nodes, we simply repeat
  the above procedure for each source node $n$ and take the disjoint union of
  the obtained path representations. The time complexity is then
  $\bigo(\size{S} \size{R})$.

  Note that we may equally well obtain $\shortestmode(R)$ by first reversing $R$
  (meaning: reversing the edge directions in $R$ and swapping $S$ and $T$),
  computing $\shortestmode$ on this reversed representation, and then reversing the
  obtained representation again. This yields a complexity
  $\bigo(\size{T} \size{R})$. 

  Assuming that path representations allow retrieving the size of $S$ and $T$ in
  $\bigo(1)$ time, we may hence first inspect which of the two is smaller, and
  then compute $\shortestmode(R)$ from the smallest of $S$ and $T$. We hence
  conclude that the final complexity is
  $\bigo(\min(\size{S},\size{T}) \size{R})$, as desired.
\inFullVersion{
    
  \medskip\uline{$\leximode(M)$} We observe that $\shortestmode(M) \subseteq \leximode(M)$. As such, we first compute, from $R$, a representation $U$ of $\shortestmode(M)$ using the method explained above. This gives us a trim representation of all pairwise-shortest paths represented by $R$. We note that $U$ is necessarily acyclic. Indeed, any cycle in $U$ can be used to generate paths $\rho$ in $U$ such that $\gamma(\rho)$ is no longer pairwise-shortest in $\mpaths(U)$, contradicting the correctness of $U$.

  We further note that, by definition of how we compute $U$, it follows that $U$ is necessarily a disjoint union of the form
  $U = U_1 \uplus U_2 \uplus \dots \uplus U_k$ where each $U_i$ is a
  representation for all pairwise-shortest paths starting from one source node
  $n_i \in S$.  Note that, again by construction, each $U_i$ has itself only one
  source node, and is of size $\bigo(\size{R})$. Let $n_i$ be the unique source
  node of $R_i$, and let $T_i$ be the set of $R_i's$ target nodes.

  We first show the reasoning for $k = 1$, i.e., when there is only one
  component in the disjoint union. For $k>1$ the result is obtained by applying
  this reasoning to each $U_i$ separately, and taking again the disjoint union
  of the results.

  Assume $k = 1$, hence $U = U_1$. We apply the following variant of the
  well-known algorithm for single-source shortest paths in directed acyclic
  graphs \cite[section 24.2]{DBLP:books/mg/CormenLRS01}. In this algorithm, we
  maintain, for every vertex $v$ of $U = U_1$, an attribute $d[v]$, which is the
  current approximation of the lexicographically-smallest sequence of edge labels that is
  used to reach $v$ from the source node $n_1$ in any path from $n_1$ to $v$ in $U$. We further maintain, in attribute $\pred(v)$, all predecessors $u$ of $v$ that can appear on any  path from $n_1$ to $v$ with edge label sequence $d[v]$.

  The algorithm is as follows.
  \begin{enumerate}
  \item Initialize, for every node $v$ in $U$, $d[v]$ to a dummy (non-existent)
    but lexicographically largest sequence and $\pred[v]$ to the empty list.
  \item Initialize $d[n_1]$ to be the empty sequence.
  \item Topologically sort the vertices of $U$.
  \item For each node $u$, taken in topologically sorted order 
    \begin{enumerate}
    \item do for each edge $e$ from $u$ 
      \begin{enumerate}
      \item let $v$ be the endpoint of $e$
      \item if $d[v]$ is lexicographically strictly larger than $d[u] \cdot \lambda(e)$ then
        \begin{itemize}
        \item set $d[v] := d[u] \cdot \lambda(e)$
        \item set $\pred[v]$ to the singleton list containing only $u$.
        \end{itemize}
        else if $d[v]$ equals $d[u] \cdot \lambda(e)$ then 
        \begin{itemize}
        \item add $u$ to $\pred[v]$.
        \end{itemize}
      \end{enumerate}
    \end{enumerate}
  \end{enumerate}

  The first node processed will necessarily be $n_1$, the unique source node of $U = U_1$. Because we are processing in topologically sorted order, whenever we process a node $v$, we have already processed all of its predecessors $u$ in $U$, and have the lexicographically smallest path that reaches $u$ starting from $n_1$.

The complexity is $\bigo(\size{V} + \size{E} \times \size{V})$ with $E$ the edges of $U_1$, and $V$ its nodes. The factor $\size{V}$ comes from the fact that the sequences $d[u]$ may be $\size{V}$ long, and for each processed edge we will need to compare such sequences. (NOte: we don't actually need to store the sequence $d[u]$, but then we would need to use the predecessor graphs to something equivalent to checking that the sequence $d[v]$ is smaller/equal than $d[u] \cdot \gamma(v)$. This is still $\bigo(\size{V})$.

This is hence $\bigo(\size{U}^2)$. For $k > 1$, we do this construction on every component, hence giving us a total complexity $\bigo(\size{U_1}^2 + \dots + \size{U_k}^2)$. Because each $U_i$ is of size $\bigo{\size{R}}$ and because $k = \bigo(\size{S})$, the total complexity is hence $\bigo(\size{S}\size{R}^2)$.

  Note that we may equally well obtain $\leximode(R)$ by first reversing $R$
  (meaning: reversing the edge directions in $R$ and swapping $S$ and $T$),
  computing $\leximode$ on this reversed representation, and then reversing the
  obtained representation again. This yields a complexity
  $\bigo(\size{T} \size{R}^2)$. 

  Assuming that path representations allow retrieving the size of $S$ and $T$ in
  $\bigo(1)$ time, we may hence first inspect which of the two is smaller, and
  then compute $\leximode(R)$ from the smallest of $S$ and $T$. We hence
  conclude that the final complexity is
  $\bigo(\min(\size{S},\size{T}) \size{R})$, as desired.}
\end{proof}

\paragraph*{Simple paths and trails.} We next turn to simple paths and trails.
We start by noting that, if P $\neq$ NP, then there does not even
exist a polynomial time algorithm for deciding if there \emph{exists} a simple path or
trail that matches a given regular expression between two given
nodes~\cite{BaganBG-pods13,MartensNT-stacs20,MendelzonW-sicomp95}. This already
implies the following:
\begin{observationrep}\label{obs:simplepaths}
  \begin{enumerate}[(a)]
  \item Given a GRPQ $\varphi = m(L)$ and graph $G$, where $m \in
    \{\simplemode,\trailmode\}$. If we can compute a \pmr (or tabular
    representation) for $\varphi(G)$ in polynomial time, then P $=$ NP.
  \item Given a \pmr $R$ for PM $M$ and $m \in
    \{\simplemode,\trailmode\}$. If we can compute a \pmr for $m(M)$  in
    polynomial time, then P $=$ NP.
  \end{enumerate}
\end{observationrep}
\begin{proof}
  \begin{enumerate}[(a)]
  \item  Assume that we can compute a \pmr $R$ for $\varphi(G)$ in polynomial time
    for every language $L$ and mode $m \in \{\simplemode,\trailmode\}$. Take $L
    = (aa)^*$. Consider the NP-complete \cite{LapaughP-networks84,BaganBG-pods13} problem of deciding if there is a simple
    path matching $(aa)^*$ from a given node $u$ to a given node $v$ in a given
    graph $G$. This problem is equivalent to deciding if there exists a path in
    $R$ from some node $x$ to some node $y$ with $\gamma_R(x) = u$ and
    $\gamma_R(y) = v$. Since the latter problem is polynomial-time solvable on
    $R$ (reachability), it would mean that we have a polynomial-time algorithm
    for the abovementioned NP-complete problem.

    The proof for $m = \trailmode$ is analogous, since deciding if there is a
    trail mathing $(aa)^*$ from $u$ to $v$ in a graph $G$ is also NP-complete \cite{MartensNT-stacs20}.
    The same arguments hold for a tabular representation of $\varphi(G)$.
    
  \item This follows from (a), as we can compute the graph product for $L$ and
    $G$ in polynomial time. The graph product is a \pmr $R$ for $\varphi'(G)$,
    where $\varphi'=L$.
  \end{enumerate}
\end{proof}
Within exponential time, however, we can even compute minimal \pmrs.
\begin{propositionrep}\label{prop:simplepaths}
  Given a \pmr $R$ for a path multiset $M$ over $G$, and selector mode $m
  \in \{\simplemode, \trailmode\}$, we can compute from $R$ a (minimal) \pmr for $m(M)$ in
  exponential time.
\end{propositionrep}
\begin{proof}
Assume that $R=(N_R,E_R,\eta_R,\gamma_R,S_R,T_R)$, and assume that $G=(N_G,E_G,\eta_G,\lambda_G)$. 
  We first consider simple paths.
  This can be done by systematically considering all paths $\rho$ in $R$ from $S_R$ to
  $T_R$ such that $\gamma_R(\rho)$ is simple. (As such, only paths of length up to
  $|N_G|$ need to be considered.) The PMR in question can be computed by taking
  the disjoint union of all these paths and minimizing the resulting \pmr.

  The procedure for trails is analogous. We now need to consider paths $\rho$
  such that $\gamma_R(\rho)$ is a trail, which means that this time we consider
  paths of length up to $|E_G|$.
\end{proof}
Although the complexity in this proposition is high, it is indeed unavoidable
by Observation~\ref{obs:simplepaths} and, furthermore, already the
worst-case \emph{size} of the set of paths represented in $R$ is exponential.

\OMIT{
It has been observed that, for the vast majority of expressions occurring in
practice, polynomial-time algorithms are possible for deciding if there is a
simple path or trail between two given nodes that matches the expression~\cite{MartensT-tods19}.
We do not dive deeply into this matter here, but study the case of $\expr =
a^*$, which is very important in practice \cite{BonifatiMT-www19}.
\begin{theoremrep}\label{theo:simple-trail-a}
  Let $G$ be a graph and $\varphi = m(a^*)$, where $m \in
  \{\simplemode,\allowbreak \trailmode\}$. We can compute a \pmr for $\varphi(G)$ in time
  polynomial in the number of rows in $\tab(\varphi(G))$.
\end{theoremrep}
\begin{proof}
  TODO. Apply Yen's algorithm.
\end{proof}
Theorem~\ref{theo:simple-trail-a} improves  is that, in
Corollary~\ref{cor:simple-trail-complexity}, it is not even possible to
guarantee time polynomial in the tabular representation $\sem{q(G)}$, since the
decision version of testing if $\sem{q(G)}$ is non-empty for general regular
languages is well-known to be \np-complete~\cite{citations}.

\wim{Not sure if we'll add the following for the first submission. Probably not.}
\begin{proposition}
  There exist queries $q = (x,p(a^+),y,m)$ with $m \in
  \{\simplemode,\trailmode\}$ and graph databases $G$ such that the smallest \pr
  for $q(G)$ is exponential in $G$.
\end{proposition}
\wim{I have an example for the above proposition, but I'm not sure if we should
  add it to the paper. It costs space and is not terribly important at this point.}
}

\paragraph*{Multiset unions.} The fundamental problem when using \pmrs for computing the
result of a union of GRPQs is to compute a path representation for the multiset
union of two \pmrs. In our setup, this is very easy to do, as it suffices to take the component-wise union of the two \pmrs.

\begin{propositionrep}
  Let $R_1$ and $R_2$ be two \pmrs. We can compute a \pmr for
  $\mpaths(R_1) \uplus \mpaths(R_2)$ in linear time
  $\bigo(\size{R_1} + \size{R_2})$.
\end{propositionrep}
\begin{proof}
  The \pr for $\mpaths(R_1) \uplus \mpaths(R_2)$ is simply given by the
  (disjoint) component-wise union $R_1\uplus R_2$  of the representations $R_1$ and $R_2$.
\end{proof}

\paragraph*{Conclusion and discussion.} It directly follows from the results in
this section (notably, Theorems~\ref{theo:product-is-a-path-representation},
~\ref{theo:singleproduct} and \ref{thm:shortest-lexi-on-prs}) that any \ugrpq
$\psi$ in which the regular languages are given as unambiguous automata and
which uses only the selector mode $\shortestmode$
can be evaluated in \emph{linear time combined complexity} when using \pmrs to
represent query outputs. This is in strong contrast to what today's systems do,
since they compute $\tab(\psi(G))$ instead, which, as illustrated in the
Introduction, is exponentially large in $\size{G}$ in general, even for $m =
\shortestmode$. We note that our linear time combined
complexity holds \emph{even when} $\psi(G)$ is infinite. Furthermore, grouping
on such \ugrpqs can also be done efficiently by Corollary \ref{cor:grouped-all},
proportional to the number of groups to be formed. The $\simplemode$ and
$\trailmode$ selector modes are more complex to evaluate on \pmrs, but we stress
that this complexity is caused by the fundamental complexity of finding simple
paths or trails that match a regular language. Also the tabular representation
faces the same complexity. 

Of course, we realize that simply representing \ugrpq outputs by means of \pmrs
may not be sufficient, and that often we want to be able to retrieve (a part of) the tabular representation, or count the number of paths retrieved. We show in the next sections that PMRs fully support this.

We conclude this section with an observation that shows that, even though
\pmr set equivalence testing is \pspace-complete in general
(Theorem~\ref{theo:prse}), it is tractable for the subclass of GRPQs specified by UFAs; any algorithm that does such a test using tabular representations
is necessarily exponentially worse.
\begin{propositionrep}\label{prop:dfa-equivalence}
  Let $\varphi_1 = \ufa_1$ and $\varphi_2 = \ufa_2$ be GRPQs and let $G$ be a
  graph database. If $R_1 = \trim(G \times \ufa_1)$ and
  $R_2 = \trim(G \times \ufa_2)$,
then we can test if $\spaths(R_1) = \spaths(R_2)$ in polynomial time.
\end{propositionrep}
\begin{proof}
The problem boils down to equivalence
  testing of UFAs. We interpret $R_1$ and $R_2$ as NFAs $A(R_1)$ and
  $A(R_2)$, just as in Appendix~\ref{app:path-mult-repr}. By
  Observation~\ref{obs:GtoNFA} the equivalence $\spaths(R_1) = \spaths(R_2)$
  holds if and only if the equivalence $\swords(A(R_1)) = \swords(A(R_2))$
  holds. We now show that $A(R_1)$ and $A(R_2)$ are both UFAs. This would conclude
  the proof, since language equivalence can be tested in polynomial time for UFAs~\cite{StearnsH-siamcomp85}.

  Assume towards a contradiction, that $A(R_i)$ with $i \in \{1,2\}$ is not a
  UFA. Then there are two runs in $A(R_i)$ that accept the same string. We
  remind that the alphabet of $A(R_i)$ is the set of edges of $G$ and every
  string accepted by $A(R_i)$ corresponds to a path in $G$. Therefore, there are
  two different paths $\rho,\rho'$ in $R_i$ that correspond to the same path
  $\gamma(\rho)=\gamma(\rho')=v_0 e_1 v_1 e_2 v_2 \cdots e_n v_n$ in $G$. Every
  node of $R_i$ is a pair $(v,q)$ with $v$ being a node in $G$ and $q$ being a
  state of $\ufa_i$. We can conclude that the first components of the nodes in
  $\rho$ and $\rho'$ are identical, i.e., they have to be $v_0,\dots,v_n$.
  However, the sequence of nodes in $\rho$ and $\rho'$ have to be different, as
  the two paths are different and by construction of $R_i$ for there cannot be
  two edges $e,e'$ in $R_i$ that have the same endpoints and refer to the same
  edge $e$ in $G$. Therefore, for at least one position in $\rho$ and $\rho'$,
  the second component of the node, i.e., the state of $\ufa_i$ differs. Thus we
  have two different runs of $\ufa_i$ that accept the same string $\lambda(e_1)
  \cdots \lambda(e_n)$. This is a contradiction to the assumption that $\ufa_i$
  is an unambiguous finite automaton and concludes the proof.
\end{proof}

\subsection{Computing Output From PMRs}\label{sec:computing-outputs-from-prs}

In this section we show that, from a given  representation $R$ of PM $M$ we can 
efficiently generate the tabular representation $\tab(M)$ when $M$ is finite, as well as compute the number of paths in $M$ or draw a random sample.

\paragraph*{Enumeration with output-linear delay.} 
We wish to be careful with what we mean by ``efficiently generate''
$\tab(M)$ from $R$ when $M$ is finite. Indeed, because $R$ can be exponentially
more succinct than $M$ and $\tab(M)$, the total time to generate $\tab(M)$ from
$R$ will obviously be exponential in $R$ in the worst case. This exponential
complexity is only due to the exponential number of tuples that we need to
generate: we will show that generating individual tuples in $\tab(M)$ from $R$
is efficient, in the sense that it takes only time proportional to the size of the tuple
being generated---independently of the size of $R$, or $M$, or $\tab(M)$. To
formalize this notion, we adopt the framework of \emph{enumeration
  algorithms}. Enumeration algorithms are an attractive way of gauging the
complexity of algorithms that need to generate large (or infinite) sets, which
have recently received significant attention in the database
community, both from a theoretical~\cite{AmarilliBMN-tods21,LosemannM-lics14,SchweikardtSV-jacm22,BerkholzGS-siglog20,Segoufin-icdt13} and practical viewpoint~\cite{IdrisUVVL-vldbj20,IdrisUV-sigmod17,TziavelisAGRY-pvldb20}.

We require the following definitions. Given an input $x$, an algorithm is said
to \emph{enumerate} a multiset $O$ if it outputs the elements of $O$ one by one
in some order $o_1, o_2, o_3, \dots$, such that the number of times an element $o$ occurs in this enumeration equals its multiplicity in $O$. (Repeated elements need not be subsequent in the enumeration.) In particular, if $O$ is a set, then the enumeration cannot contain duplicates. It enumerates $O$
\emph{with output-linear delay} if the time required to output the $i$-th element $o_i$,
measured as the difference in time between outputting $o_{i-1}$ (or the start of
the algorithm, when $i=1$) and finishing outputting $o_i$, is proportional to
the size of $o_i$, independent of the size of $O$ or of the input $x$. If $O$ is
finite, then it is also required that the algorithm terminates immediately after
outputting the last element.  In that case, the total time that the algorithm
takes to enumerate $O$ is hence $\bigo(\size{O})$, i.e., linear in $O$.

\begin{propositionrep}\leavevmode
  \label{prop:enumeration-delay}
  Let $M$ be  a finite path multiset on a graph $G$.
  \begin{enumerate}
  \item From a trim PMR $R$ of $M$ we can enumerate both $M$ and $\tab(M)$ with output-linear delay.
  \item From a trim grouped PMR $R$ of $\group{S}(M)$ we can enumerate both
    $\group{S}(M)$ and $\tab(\group{S}(M))$ with output-linear delay, for any
    non-empty $S \subseteq\{\src,\tgt\}$.
  \end{enumerate}
\end{propositionrep}
\begin{proof}
  (1)   We focus on enumerating $\tab(M)$; the other cases are
  similar.  Note that to enumerate $\tab(M)$ it suffices to enumerate the paths
  $\rho$ from $S_R$ to $T_R$ in $R$: every such path corresponds to a precisely
  one tuple $(\src(\gamma_R(\rho), \tgt(\gamma_R(\rho), \gamma_R(\rho))$ in
  $\tab(M)$. (Paths $\rho$ that have the same $\gamma_R(\rho)$ will cause this
  tuple to be output multiple times, as desired.) Furthermore, because $R$ is
  trim every node in $R$ occurs on some path from $S_R$ to $T_R$. We can hence
  enumerate the paths from $S_R$ to $T_R$ in $R$ by means of a variant of
  depth-first search (DFS): start from a node $v_0 \in S_R$, and perform DFS
  until we reach a node $v_k \in T_R$ (which we are guaranteed to). During the
  search, we incrementally build the the path $\rho$ being traversed. Once
  completed, we can output the corresponding tuple, and backtrack to the last
  node in $\rho$ that has at least one edge we did not traverse yet (given the
  current start node $v_0$); we then continue DFS from that node again to
  generate the next path. Once we backtrack all the way to $v_0$ we pick another
  source node, and start over. It is not difficult to see that the time required
  to build a complete path $\rho$ is linearly proportional to $\size{\rho}$,
  which is proportional to tuple to be output. The only technical detail to
  resolve is the fact that a found path $\rho$ may be significantly shorter than
  the next path $\rho'$. Because we also need to backtrack from $\rho$ to reach
  $\rho'$, strictly speaking the delay for $\rho'$ is not output-linear. This
  can be solved, however, by performing the backtracking on $\rho$ before
  outputting the tuple corresponding to $\rho$, so that the backtracking time of
  $\rho$ is put at $\rho$'s expense.

  (2) Follows from the fact that, since each  \pmr $R_i$ in the trim grouped \pmr $R$ of $\group{S}(M)$ is itself trim, we can compute $\mpaths(R_i)$ in time $\bigo(\size{\mpaths(R_i)})$ by (1). The enumeration algorithm then simply consists of iterating over the \pmrs in $R$, and computing $\mpaths(R_i)$ for every such \pmr. (Possibly adding $\src(\mpaths(R_i))$ and/or $\tgt(\mpaths(R_i))$ in case of $\tab(\group{S}(M))$.)
\end{proof}

Coupled with the results of Subsection~\ref{sec:rpq-evaluation},
Proposition~\ref{prop:enumeration-delay} tells us that we can evaluate an UGRPQ
$\varphi$, which uses either the $\shortestmode$ selector, or no selector at
all, on a graph database $G$, by running a polynomial preprocessing phase for
computing the \pmr for $\varphi(G)$, and then enumerating the results one-by-one
in time that is proportional to the length of the output path. In a sense, one
could argue that such enumeration is optimal, since this is the time it takes to
write own the output.
For $\trailmode$ and
$\simplemode$ the same guarantee on enumerationi holds, but
constructing the appropriate \pmrs now requires an exponential preprocessing
phase.

We conclude this subsection by observing how \pmrs can be used to count the
number of query results, or sample paths in a GRPQ output uniformly at random.
The former kind of result is relevant for dealing with queries that involve
projection and joins (Section~\ref{sec:gcrpqs}) and the latter can be useful to
provide uniform sampling guarantees to GQL's ANY-mode~\cite{GQL-industry}, if desired.

\newcommand{\unarysize}[1]{\|#1\|}
\begin{propositionrep}\label{prop:counting}
  Let $R$ be a trim PMR.  Then we can
  \begin{enumerate}
  \item count the number of paths in $\mpaths(R)$ in linear time,
    where the returned result is $+\infty$ if $\mpaths(R)$ is infinite;
  \item if $\mpaths(R)$ is a finite multiset, uniformly at random sample a path
    in $\mpaths(R)$ in linear time.
  \item given a natural number $n \in \nat$, uniformly at random sample
    a path from the submultiset of all paths of length $n$ in $\mpaths(R)$, in 
    time
$\bigo(n \size{R})$.
  \end{enumerate}
\end{propositionrep}
\begin{proof}
  (1) First check whether $\mpaths(R)$ is infinite, which is the case if and
  only if $R$ is acyclic. Acyclicity of $R$ can be checked in linear time by
  depth first search. If $R$ is cyclic, return $+\infty$. Otherwise, the number
  $\size{\mpaths(R)}$ that we need to return equals the number of paths from
  $S_R$ to $T_R$ in acyclic graph $R$. This is well-known to be computable in
  linear time by means of a form of dynamic programming:
  \newcommand{\cnt}{\textsf{cnt}}
  \begin{itemize}
  \item First, compute a reverse topological sort $u_1, \dots, u_k$ of the nodes in $R$. Hence, if there is an edge from $v$ to $w$ in $R$, $w$  will occur before $v$ in the topological sort. 
  \item For $i = 1$ to $k$:
    \begin{itemize}
    \item If $u_i$ is a terminal node (meaning that it is necessarily in $T_R$ because of trimedness), set $\cnt(u_i) := 1$.
    \item If $u_i$ has outgoing edges, then set \[ \cnt(u_i) := \sum_{e \in E_R, \eta_R(e) = (u_i,v)} \cnt(v).\] Note that, because of the reverse topological sort, in the last step all successors of $u_i$ have already been processed, and hence have a $\cnt$ score computed. 
    \end{itemize}
    Observe that $\cnt(v)$ hence computed is the number of paths from $v$ to a node in $T_R$.
  \item Finally, return $\sum_{v \in S_R} \cnt(v)$
  \end{itemize}
  The topological sort can be computed in linear time. After that, we traverse each node and each edge once. By storing the $\cnt$-values in a hash map we can also retrieve the count-value of a node in $\bigo(1)$ time in the RAM model. Overall, the algorithm is hence in linear time.
  
(2) This follows from the construction for (1), since we can store, for each
node $u$ in $R$ the number $k_u$ of paths in $R$ that start from $u$. Since $R$ is
trim, every such path corresponds to a single element in $\mpaths(R)$. Hence, we can
uniformly at random sample a path by traversing $R$ from beginning to end and,
at each node $u$ with out-neighbours $u_1,\ldots,u_n$, choose an outgoing edge
according the distribution of $k_{u_1},\ldots,k_{u_n}$.

\smallskip
  (3) First compute the total number $N$ of paths in $\mpaths(R)$ in linear time, using the reasoning outlined in (1) above. This produces, for every node $v$, the number of paths $\cnt(v)$ leading from $v$ to a node in $T_R$. Because $R$ is trim, every node in $R$ participates on some path from a node in $S_R$ to a node in $T_R$. 
  We can then sample uninformly at random as follows. Define the following probability distribution $P(u)$ on nodes in $S_T$: $P_{\src}(u) = \frac{\cnt(u)}{N}$. Furthermore, define the conditional probability $P_{\textsf{edge}}(e \mid v)$ of an edge $e$ given an node $v$ by
\[ P_{\textsf{edge}}(e \mid u) =
  \begin{cases}
    \frac{\cnt(v)}{\cnt(u)} & \textsf{if $\eta_R(e) = (u,v)$} \\
    0 & \textsf{if $e$ does not have $u$ as source.}
  \end{cases}
\]

Note that both probabilities can be computed in constant time, given $u$ and $v$. We can then sample a path from $R$ uniformly at random by first sampling a node in $S_R$ according to $P_{\src}$ to determine an initial node, and then, as long as we have not yet reached a node in $T_R$, draw an edge incident on the current node according to $P_{\textsf{edge}}$. Because $\mpaths(R)$ is finite, $R$ must be acyclic, and so this process necessarily terminates. In particular, because of trimedness, it necessarily terminates in a node in $T_R$.

The probability of generating the path $u_0 e_1 v_1 \dots e_k v_k$ is then
\begin{align*}
  & P_{\src}(u_0) P_{\textsf{edge}}(e_1 \mid u_0) P_{\textsf{edge}}(e_2 \mid u_1) \cdots P_{\textsf{edge}}(e_k \mid u_{k-1}) \\
  & = \frac{\cnt(u_0)}{N} \frac{\cnt(u_1)}{\cnt(u_0)} \frac{\cnt(u_2)}{\cnt(u_1)} \dots \frac{\cnt(u_k)}{\cnt(u_{k-1})} \\
  & = \frac{\cnt(u_k)}{N}\\
    & = \frac{1}{N}
\end{align*}
where the last equality follows because $\cnt(u_k) = 1$ for nodes in $T_R$ due to acyclicity.

(3) (Sketch.) Construct a \ufa $A$ that recognizes all paths of length $n$, in $\bigo(\unarysize{n})$  time. Construct the product of $R$ with $A$ in time $\bigo(\unarysize{n} \size{R})$ and trim it, in the same time. The resulting PMR represents the submultiset of all paths of length $n$ in $\mpaths(R)$. Because there can be at most a finite number of paths of a given length in a graph, and because $\mpaths(R)$ are all paths in the same graph, the submultiset is necessarily finite. Then, apply (2) to obtain the desired result.
\end{proof}

\subsection{Graph Projections}
\label{sec:graphprojections-complexity}
Let $\psi$ be a \ugrpq and let $G$ be a graph database.
While all systems that we know of compute the tabular representation
$\tab(\psi(G))$ for presentation to the user, many systems also support
presenting the user with a \emph{graph projection} of $\psi(G)$, which is the
subgraph of $G$ that consists of the nodes and edges that are used in some path
in $\psi(G)$.

Formally, the \textit{graph projection} of a path multiset $M$ on a graph $G$,  denoted by
$\semg{M}$, is the subgraph $G'$ of $G$
consisting of:
\begin{itemize}
\item $N_{G'} := \bigcup_{\rho \in M} \textsf{nodes}(\rho)$;
\item $E_{G'} := \bigcup_{\rho \in M} \textsf{edges}(\rho)$;
\item $\eta_{G'}(e):=\eta_G(e)$, for $e\in E_{G'}$;
\item $\lambda_{G'}(e) := \lambda_G(e)$, for $e\in E_{G'}$.
\end{itemize}
In this definition, for a path $\rho= v_0 e_1 v_1 e_2 v_2 \cdots e_n v_n$, we
write $\textsf{nodes}(\rho)$ to denote the set $\{v_0,v_1,\ldots ,v_n\}$, and
$\textsf{edges}(\rho)$ to denote the set $\{e_1,\ldots ,e_n\}$.

Graph projections are hence themselves graphs, like \pmrs. And, like \pmrs, they
can be exponentially more succinct than $\psi(G)$. Indeed, consider the graph of
Figure~\ref{fig:2n-shortest-in-graph} and assume that $\psi$ returns all paths
from $x$ to $y$. The graph projection of $\psi(G)$ is simply $G$
itself. We stress, however, that, unlike \pmrs, graph projections are not
always lossless, which we already argued in the Introduction. As another
example, consider the path multiset of all cycles in
the graph of Figure~\ref{fig:propertygraph} of even length  from Mike to
Mike. A (lossless) PMR for this path multiset is shown in Figure~\ref{fig:mike-mike-even}. By contrast, the graph projection of this path multiset is simply the subgraph of Figure~\ref{fig:mike-mike-even} consisting of the nodes $\{\nodeid{a1}, \nodeid{a3}, \nodeid{a5}\}$ and edges $\{\edgeid{t1},\edgeid{t7},\edgeid{t8}\}$. This subgraph has a cycle of length $3$, which is \emph{not} in the PM that we want to represent.

Today, graph query engines such as Cypher primarily return a visualization of
the graph projection of $\psi(G)$ \emph{if they can}. Since the graph
projection of $\psi(G)$ can be exponentially more succinct than $\psi(G)$ and
$\tab(\psi(G))$ itself, this raises the question in which cases one can avoid
explicitly computing $\psi(G)$ or $\tab(\psi(G))$, which means avoiding an intermediate
(exponential) step.  To the best of our knowledge, Cypher currently always
computes $\tab(\psi(G))$ first, since they use it as the input for computing
$\semg{\psi(G)}$.

The graph projection of $\psi(G)$ can be immediately computed from any \pr of
$\psi(G)$. In fact, if we have a \pmr $R$ of $\psi(G)$, then the graph
projection of $\psi(G)$ is just the \emph{image of $R$ in $G$ under the
  homomorphism $\gamma_R$}, defined to be the subgraph $G'$ of $G$ that consists
of $N_{G'} := \{\gamma_R(v) \mid v \in N_R\}$, $E_{G'} := \{\gamma_R(e) \mid e
\in E_R\}$, $\eta_{G'} = \restr{\eta_G}{E_{G'}}$, and $\lambda_{G'} =
\restr{\lambda_G}{E_{G'}}$.

\begin{propositionrep}
  Let $R$ be a trim \pmr on a graph $G$. Then the image of $R$ equals the graph
  projection $\semg{\mpaths(R)}$.
\end{propositionrep}

Since computing the graph projection of a path representation is clearly
possible in linear time, we immediately the following.
\begin{corollary}
  Let $\psi$ be a \ugrpq and $G$ a graph, then we can compute the graph projection $\semg{\psi(G))}$ in the same time as computing a PMR for $\psi(G)$.
\end{corollary}

 \section{Conjunctive GRPQs}
\label{sec:gcrpqs}

We now consider \emph{conjunctive generalized regular path queries (CGRPQs)},
which extend GRPQs with joins. A CGRPQ is an expression of the form \[Q =
  (z_1,\varphi_1,z'_1), \ldots, (z_k,\varphi_k,z'_k)\;,\] where $\varphi_i$ is a
GRPQ for each $i \in [k]$. We define the semantics of CGRPQs immediately in
terms of their tabular output, which is in line with GQL pattern matching
queries~\cite{GQL-industry}. To this end, let $\bar x = (x_1,\ldots,x_\ell)$ and
$\bar y = (y_1,\ldots,y_\ell)$ be tuples of the same arity. We denote by $\bar x
\succcurlyeq \bar y$ that equalities in $\bar x$ should also hold in $\bar y$,
that is, if $x_i = x_j$ then $y_i = y_j$ for all $i \in [\ell]$. The (tabular)
output of $Q$ on graph $G$ is denoted by $\tab(Q(G))$ and is defined as
\begin{multline*}
  \multileft (u_1, \rho_1, v_1, \ldots, u_k, \rho_k, v_k) \mid \rho_i \in
  \varphi_i(G), u_i = \src(\rho_i), v_i = \tgt(\rho_i), \\
  (z_1,z'_1,\ldots,z_k,z'_k) \succcurlyeq (u_1,v_i,\ldots,u_k,v_k) \multiright\;.
\end{multline*}
Notice that, again, this multiset is a set.

\paragraph*{Using Grouped Path Representations.}
The output of a CGRPQ is therefore a table in which each row contains $2k$ nodes
and $k$ paths. Similarly to Section~\ref{sec:grpqs}, we can now use grouped path
multisets (GPMs) and grouped \pmrs to succinctly represent
this output.

Fix $Q$ and $G$. We define $E = \{(u_1,v_1,\ldots,u_k,v_k) \mid (u_1, \rho_1,
v_1, \ldots, \allowbreak u_k, \allowbreak \rho_k, v_k) \in \tab(Q(G))\}$ as the set of endpoint tuples
in $Q(G)$. We define, for each $i \in [k]$, the set $M_i = \{\rho_i \mid (u_1,
\rho_1, v_1, \ldots, u_k, \rho_k, v_k) \in \tab(Q(G))\}$ as the set of paths
that participate in the answers to the $i$-th GRPQ in $Q$ on $G$. A \emph{fully
  grouped table representation of $\tab(Q(G))$} is the set of tuples
\begin{multline*}
  \Gamma(Q(G)) = \multileft (u_1, M'_1, v_1, \ldots, u_k, M'_k, v_k) \mid (u_1,v_1,\ldots,u_k,v_k) \in E,\\
  M'_i = \sigma_{\{u_i\},\{v_i\}}(M_i) \text{ for all } i \in [k]  \multiright\;.
\end{multline*}
Notice that, for each $i \in [k]$, the partition ${\mathcal H}_i = \{M'_{i} \mid
(u_1, M'_1, v_1, \allowbreak \ldots, \allowbreak u_k, M'_k, v_k) \in
\Gamma(Q(G))\}$ is a pairwise grouped path multiset. We will use a grouped \pmr
$S_i$ to represent ${\mathcal H}_i$. Notice that the number of multisets in
${\mathcal H}_i$ is quadratic, since the number of groups for $i \in [k]$ is
determined by the number of node pairs $(u_i,v_i)$. So, even though the fully
grouped table representation $\Gamma(Q(G))$ may still contain exponentially many
tuples (due to the exponentially many elements in $E$), we can represent all the
different path multisets in $\Gamma(Q(G))$ using quadratically many \pmrs in the
worst case.

\paragraph{Evaluating Chain Queries using \pmrs.}
We discuss how \prs are helpful for evaluating CGRPQs, but limit ourselves to
\emph{chain queries} for space reasons. A \emph{chain query} is a CGRPQ of the
form
\[Q = (z_1,L_1,z'_1), \ldots, (z_k,L_k,z'_k)\;,\] where $z'_i = z_{i+1}$ for all
$i \in [k-1]$ and all variables are pairwise different otherwise. We only
consider GRPQs of the form $L_i$ here and compute \pmrs for all paths in
$L_i(G)$ that are useful for results in $Q(G)$. Our results in
Section~\ref{sec:computing-prs} show how to use these to obtain \pmrs for more
complex GRPQs such as $\shortestmode(L_i)$, etc.

In particular, we show that we can compute all the grouped \pmrs for the path
multisets in $\Gamma(Q(G))$ in cubic time combined complexity if the languages
$L_i$ are given as UFAs. Since there can be quadratically many path
multisets in the output, this gives us linear time per path multiset.
Furthermore, since the number of tuples in $\Gamma(Q(G))$ is in $\Theta(|G|^k)$,
this shows that computing \pmrs for representing the paths in $Q(G)$ is not the
major bottleneck.\footnote{Observe, however, that the \emph{number of tuples} in
  $Q(G)$ can be in $\Omega(2^{k|G|})$, even when only the $\shortestmode$ mode
  is used.}

\begin{enumerate}[(1)]
\item Let $\ufa_i$ be a UFA for each language $L_i$. We can assume
  \mbox{w.l.o.g.} that each $\ufa_i$ has a single initial and final state. Let
  \ufa be a UFA the language $L_1 \cdots L_k$, obtained by concatenating the $\ufa_i$. 
  Let $q_0,\ldots,q_k$ be the start states of the sub-UFAs for $L_1,\ldots,L_k$.
\item Using Theorem~\ref{theo:product-is-a-path-representation}, we can compute a
  trim \pmr $R$ for $\ufa(G)$ in time $\bigo(|Q||G|)$. For each $i \in \{0,\ldots,k-1\}$, let
  $U_i = \{\gamma(u,q_i) \mid u \in N_G\}$ and let $U_k = \gamma(T_R)$. Notice that $U_0 = \gamma(S_R)$.
\item For every $i \in [k]$, compute a \pmr $R_i$ for
  $\sigma_{U_{i-1},U_i}(L_i(G))$. This takes time $\bigo(k|Q||G|)$.
\item Compute grouped \pmrs for $\group{\src,\tgt}(R_i)$. Using the notation in
  Corollary~\ref{cor:grouped-all}, this takes time $\bigo(k|U_{i-1}U_i||R_i|)$. 
\end{enumerate}
Then, for each $i \in [k]$, the grouped \pmrs in step (4) represent the
multisets $\{M'_{i} \mid (u_1, M'_1, v_1, \ldots, u_k, M'_k, v_k) \in
\Gamma(Q(G))\}$ and took time $\bigo(k|Q||G|^3)$ to compute. This time bound may
seem high, but it only occurs if the sizes of the sets $U_i$ are $\Theta(|G|)$,
which is unlikely in practice. Furthermore, computing each path separately,
which is the current state of the art, is exponentially worse. Finally, the
above procedure was written to show the potential of using \pmrs. It still has
significant room for optimization, for instance, the representations $R_i$ in
(3) are actually substructures of $R$ in (2) and do not have to be computed.

\paragraph{A Faster Algorithm in the Presence of Projection.}
Next we show that \pmrs can even bring the run-time down from exponential to
linear if the query is unary. To this end, consider queries of the form 
\[\pi_1(Q)\,,\] where $Q$ is a chain query as defined before. (For
simplicity, we simply project on the first variable of the query. This is,
however, not essential for the argument.) Such queries are very natural. For
instance, {\small
\begin{verbatim}
    MATCH (x) WHERE
    (x:Person) -[:Job]-> (y) -[:SubClassOf*]-> (z:Artist)
\end{verbatim}
}
\noindent is a query of the required form, which returns persons whose
occupation is a subclass of
``Artist''. We define
\begin{align*}
  \pi_1(Q)(G) & = \bag{u \mid \exists (u,\rho_1,v_1,\ldots,u_k,\rho_k,v_k) \in \tab(Q)(G)}\;,
\end{align*}
where each node $u$ occurs in the answer as often as there are tuples
$(u,\rho_1,v_1,\ldots,u_k,\rho_k,v_k)$ in $\tab(Q)(G)$.

For computing $\pi_1(Q)(G)$, we proceed as follows. We repeat steps (1) and (2)
as in the evaluation algorithm for chain queries and proceed as follows.
\begin{enumerate}
\item[(3)] For every node $u$ in $U_0$, let $\ell_u$ be the number of paths in
  $R$ that go through nodes of the form $(u,q)$. These can be computed in linear
  time using the techniques for Proposition~\ref{prop:counting}. (Essentially,
  the task is, given the \pmr $R$ and disjoint sets of nodes
  $\{V_1,\ldots,V_n\}$, to count the number $\ell_i$ of paths in $R$ that start
  in $S_R$ go through any node in $V_i$ and end in $T_R$, for every $i \in [n]$.
  This can be done by a single dynamic programming algorithm in linear time. The
  algorithm essentially computes, for every node in $R$, the number of paths leading
  to $T_R$ and the number of paths coming from $S$.)
\item[(4)] Output the multiset $M$ such that $M(u) = \ell_u$ for every such node
  $u$.
\end{enumerate}

Since all the steps in the above procedure are linear, we obtain:
\begin{proposition}
  For a unary chain query $\pi_{1}(Q)$ and graph $G$, the answer $\pi_1(Q(G))$
  can be computed in linear time combined complexity $\bigo(|Q||G|)$.
\end{proposition}

 \section{Related Work}
\label{sec:related-work}

Queries over graph-structured data have been extensively studied,
e.g.~\cite{AbiteboulQMWW-jodl97,CalvaneseGLV-kr00,ConsensM-pods90,FlorescuLS-pods98,MendelzonW-sicomp95}.
A popular means of querying are conjunctive regular path queries
(CRPQs)~\cite{CalvaneseGLV-kr00, DeutschT-dbpl01, FlorescuLS-pods98,
  FigueiraGKMNT-kr20}, which return tuples of nodes which are connected in a way
predefined by the CRPQ. This mode of evaluating (conjunctive) regular path
queries has dominated the research landscape for decades \cite{Barcelo-pods13}
and is also the mode of evaluation for regular path queries in SPARQL
\cite{sparql11,LosemannM-tods13,ArenasCP-www12}.

However, as the data gets larger and more complex, it gets more and more
important to include paths in the output of the query~\cite{KochutJ-esws07}.
Indeed, G-Core~\cite{gcore}, a result of intense collaboration between industry
and academia, proposes to treat paths first-class citizens in graph databases
and, hence, allows queries to return them. GQL~\cite{GQL-industry}, the upcoming
ISO standard for querying property graphs, builds on the
G-Core proposal, but takes a perspective closer to industry. The industry/academia collaboration for G-Core and GQL takes place
under the auspices of the LDBC, which also generated work on keys for property graphs~\cite{pg-keys} and threshold queries~\cite{threshold-pvldb22}.

The two lines of work that are the most closely connected to ours are the following.

\paragraph*{Factorized databases.} Olteanu and co-authors have proposed Factorized Database Representations (FDBs)~\cite{DBLP:journals/pvldb/BakibayevKOZ13,DBLP:journals/pvldb/BakibayevOZ12,DBLP:journals/tods/OlteanuZ15,DBLP:journals/pvldb/Olteanu20} as a means of succinctly representing query results, possibly exponentially more succinct than traditional tables, while still allowing enumeration of such tables with constant delay. Factorized databases hence share important properties with the path multiset representation proposed here. We stress, however, that FDBs and PMRs are incomparable.  Indeed, on the one hand PMRs are more expressive than FDBs: FDBs were developed to represent results of traditional conjunctive queries on relational databases (or, more generally, relational algebra queries), not for representing results of GRPQs applied to graphs. In particular, conjunctive queries, when evaluated  on graphs, can only return paths whose length is bounded by the number of atoms in the query. By contrast, GRPQs can return paths of unbounded length. Consequently FDBs can only represent paths of bounded length, while PMRs can represent paths of unbounded length.

On the other hand, FDBs are more expressive than PMRs. This is because FDBs can represent results of any conjunctive query, and, on graphs, conjunctive queries can express patterns such as triangles that do not adhere to a path topology. While FDBs can represent such expressive graph patterns, \pmrs are limited to paths.

Finally, FDBs and PMRs are fundamentally distinct mathematical objects. FDBs represent relational tables as an expression involving unions and Cartesian products, whereas PMRs are graphs, endowed with a homomorphism.

\paragraph{Finite state automata and ECRPQs.}
Some of our constructions (notably Definition~\ref{def:graph-product}) are
heavily inspired on the product construction for non-deterministic finite
automata~\cite{HopcroftUIntroduction07}. Indeed, taking the ``product'' of a
graph and an NFA is a folklore method for computing the output of regular path
queries in the literature. Barcel\'o et al.~\cite{BarceloLLW-tods12} used a
different but similar construction to investigate query evaluation for
\emph{extended conjunctive regular path queries (ECRPQs)} which, as us, also
extend CRPQs with the ability to include paths in the output of the query, but
also to define complex semantic relationships between paths, using regular
relations. Like us, they provide an automaton construction that can represent
both nodes and paths in the output. The remainder of their work is quite
different from ours, since they had a different focus. They provided a picture
of what can be implemented in standard query languages in terms of complexity,
including concerning questions such as query containment. To deal with relations
on paths, they define a notion of convolutions of graph databases and queries,
that reduces the evaluation of ECRPQs to the evaluation of CRPQs. Our focus on
compact representations for query evaluation, and their interaction with
modular operations in query plans, is therefore quite different.

 \section{Conclusions}
\label{sec:conclusions}

We presented the concept of \emph{path multiset representations (\pmrs)}, which
allow to represent multisets of paths in an exponentially succinct manner. We
believe that such a concept is necessary for ensuring returning paths
in modern graph query engines remains a tractable endeavor. Indeed, while
the number of (even shortest) paths that match regular path queries can become
prohibitively large, \pmrs allow to represent these using linear space in terms
of combined complexity.

This paper presented a wide number of results that involve the incorporation of
\pmrs in query language engines, using a modular query evaluation approach,
which is typical for how database query engines work. By showing how \pmrs hold
up when considering grouping operators, unions, joins,
projection, counting, and random sampling, we have gone significantly beyond
the restricted setting that is typically  considered in research, i.e.,
regular path queries and set semantics.

We note that \pmrs may even be useful in terms of query language design. An
important reason why selectors and restrictors to finite sets of paths are used
in in modern graph query languages \cite{cypher,GQL-industry} is because the
community does not know how to deal with infinite sets of paths. But such
restrictions can be detrimental to query languages. For instance, by
restricting ourselves to data structures that can only represent finite sets of
paths, we intuitively make logical and physical operators less composable, which
in turn may rule out operations further in the query plan. For example, it is
not possible to randomly sample a path of length $n$ between two nodes, if we
have discarded the paths of this length in a previous computation step.
\pmrs, however, can represent the infinite sets that are returned by regular
path queries in a finite manner, as Example~\ref{ex:state-information} and
Theorem~\ref{theo:product-is-a-path-representation} illustrate. It is therefore an
interesting question whether a composable algebra for graph querying that allows infinite
intermediate results can be built up using \pmrs or a variation thereof.

\section*{Acknowledgments}
We are grateful to Matthias Hofer for valuable discussions and to Wojciech
Czerwi\'nski for pointing us to \cite{Tzeng-ipl96}. This work was
supported by the ANR project EQUUS ANR-19-CE48-0019; funded by the Deutsche
Forschungsgemeinschaft (DFG, German Research Foundation) -- project number
431183758.
Vansummeren was supported by the Bijzonder
Onderzoeksfonds (BOF) of Hasselt University (Belgium) under Grant No.
BOF20ZAP02. Vrgo\v{c} was supported by ANID -- Millennium Science Initiative
Program -- Code ICN17\_002.

\bibliographystyle{ACM-Reference-Format}
\bibliography{references}

\appendix

\end{document}